\documentclass[twocolumn,showpacs,preprintnumbers,amsmath,amssymb]{revtex4}

\usepackage{epsfig,psfrag}
\usepackage{dcolumn}
\usepackage{bm}
\usepackage{graphicx}
\usepackage{color}

\begin{document}


\title{Towards full counting statistics for the Anderson impurity model}

\author{A.~O.~Gogolin$^1$ and A.~Komnik$^{2,3}$}

\affiliation{${}^{1}$ Department of Mathematics, Imperial College London,
180 Queen's Gate, London SW7 2AZ, United Kingdom \\
${}^2$~Service de Physique Th\'eorique, CEA Saclay, F--91191
Gif-sur-Yvette, France \\ ${}^3$~Physikalisches Institut,
Albert-Ludwigs-Universit\"at,
  D--79104 Freiburg, Germany
}

\date{\today}

\begin{abstract}
We analyse the full counting statistics (FCS) of the charge
transport through the Anderson impurity model (AIM) and similar
systems with a single conducting channel.
The object of principal interest is the
generating function for the cumulants of charge current
distribution. We derive an \emph{exact analytic formula} relating
the FCS generating function to the self energy of the system in
the presence of the measuring field. We first check that our
approach reproduces correctly known results in simple limits, like
the FCS of the resonant level system (AIM without Coulomb
interaction). We then proceed to study the FCS for the AIM both
perturbatively in the Coulomb interaction and in the Kondo regime
at the Toulouse point (we also study a related model of a spinless
single-site quantum dot coupled to two half-infinite metallic
leads in the Luttinger liquid phase at a special interaction
strength). At zero temperature the FCS turns out to be binomial
for small voltages. For the generic case of arbitrary energy
scales the FCS is shown to be captured very well by
generalisations of the Levitov-Lesovik type formula. Surprisingly,
the FCS for the AIM indicates a presence of coherent electron pair
tunnelling in addition to conventional single-particle processes.
By means of perturbative expansions around the Toulouse point we
succeeded in showing the universality of the binomial FCS at zero
temperature in linear response. Based on our general formula for
the FCS we then argue for a more general \emph{binomial theorem}
stating that the linear response zero-temperature FCS for
\emph{any interacting} single-channel set-up is always binomial.

\end{abstract}


\maketitle

\section{Introduction}     \label{introduction}
The Anderson impurity model is one of the best studied models in
condensed matter theory \cite{andersonspin,hewson}. Despite being
exactly solvable by means of the Bethe Ansatz (BA) method in the
wide range of equilibrium parameters
\cite{kawakamiokiji,vigmantsvelik,tsvelickwiegmann}, its
non-equilibrium properties are not yet fully understood. Notable
exceptions are the works on the non-linear $I-V$ characteristics
\cite{KSL,KSLPRL,SH}. It has first been realised by Schottky
\cite{schottky}, that the current autocorrelation spectrum
(sometimes also called noise) carries information about the charge
of particles participating in transport. The investigation of
these properties has been started recently
\cite{ding,meirgolub,hamasaki}. However, the current-voltage
characteristics and noise spectra are only the lowest order
moments of the full current distribution function, which is needed
to completely characterise the transport properties of the system.
Although it is still quite challenging to access even the noise
correlations in experiments, in recent years it became possible to
measure the third irreducible moment (third cumulant) of the
current distribution function \cite{reulet}. It turned out to
carry information about the influence of the electromagnetic
environment on the transport through the system under
consideration \cite{kindnaza}. Moreover, it has was been argued
that the third cumulant is more suited for measuring the charge of
current carrying excitations than the noise correlations
\cite{reznikov}. In order to meet future experimental needs it is
therefore natural to analyse the full counting statistics of the
AIM.

The principal question is this: what are the effects of the
electron--electron interactions on the FCS? Is it possible to gain
insight into the properties of a strongly correlated electron
system by studying its FCS distribution function? We provide at
least a partial answer in this paper. The answer turns out to be
on the negative side, though it is a constructive one: we find
that the binomial statistics is {\it universal} in the low
temperature linear response limit. (Interactions only affect the
magnitude of the effective transmission coefficient.) At high
voltage (temperature) the effects of the interactions are indeed
profound (see main text) if more model dependent.

The AIM model is characterised by a number of different
parameters: the electronic tunnelling amplitude $\gamma$
between of the impurity  level (which we also shall sometimes
call `dot' later)  and the external electrodes,
its energy $\Delta_0$, and the strength of the
Coulomb interaction on the dot $U$. There are three
different transport regimes: (i) resonant level case, when $U$ is
vanishingly small in comparison to all other energy scales; (ii)
Kondo dot regime, when $U$ is large and when the dot level
lies deep below the Fermi energies in the electrodes; (iii)
mixed valence regime, which comprises all other possibilities. The
most interesting situation is (ii) when the dot is permanently
populated by a single electron. The so-called Kondo-resonance
(also known as Abrikosov-Suhl resonance) in the local density of
states leads to a significant increase of conductivity, which has
recently been observed in experiments on ultra-small quantum dots
\cite{goldhaber-gordon,kouwenhoven}. This phenomenon is a
signature of the Kondo effect and is a result of the exchange
interaction between the local spin degree of freedom on the dot
with those in the leads. One important feature of this effect is
the fact that it is growing stronger as the temperature is
lowered. From the mathematical point of view that means that the
exchange interaction is a relevant operator in the renormalisation
group (RG) sense, resulting in a new ground state where the
local spin is absorbed, the leads are coherent and the conductance
is maximal (perfect).

The equilibrium Kondo model, being one special case of AIM model,
is integrable by means of the BA technique
\cite{tsvelickwiegmann,andreifuruyalowenstein}. While it is
possible to infer the non-linear $I-V$ from the knowledge of
scattering matrix, it is not yet clear whether an extraction of
noise or any other higher order correlations of current is
feasible. It has been pointed out by Toulouse \cite{Toulouse},
that the Kondo model allows a trivial diagonalisation for one
special parameter constellation, when the whole Hamiltonian
becomes quadratic in fermion fields. In addition to this very
useful feature the Kondo model at the Toulouse point turns out to
be representative for the low energy behaviour of a generic Kondo
model \cite{book}, reproducing all essential details of the latter
in the low energy sector. For the Kondo dot a similar procedure
has been developed by Emery and Kivelson in \cite{EK} and refined
by Schiller and Hershfield \cite{SH} in order to access the
non-equilibrium current-voltage as well as noise properties. As
has been shown in \cite{long}, this approach can be applied to
access the FCS as well.

Contrary to the single-channel Kondo model, which maps on the
conventional non-interacting resonant level (RL) model at the
Toulouse point, the Kondo dot under the same conditions is
described by the Majorana RL Hamiltonian \cite{book}. It has been
demonstrated in \cite{ourPRL,ourPRB}, that an RL coupled to two
half-infinite Luttinger liquids (LL) at the special interaction
parameter $g=1/2$ can be re-written in terms of a Majorana RL
model as well. In this way one can obtain the exact FCS for a
genuine interacting system. Related systems have been analysed
before, see \cite{AM,KT} (`Coulomb blockade' dots). Below we
investigate how the two models are related.

The outline of the paper is as follows. In Section \ref{SectionI}
we present a further development of the Levitov-Reznikov
\cite{reznikov} approach to the FCS calculation in tunnelling
set-ups. In order to test our Hamiltonian formalism we perform an
explicit calculation of the generating function for the FCS of a
simple tunnelling contact between two metallic electrodes, see
Section \ref{tunnjunction}. Section \ref{FCSAI} starts with a
reproduction of the known FCS for the RL model, which is the
spin-less version the of AIM set-up without Coulomb interaction.
Next we derive Eq.~(\protect\ref{fi}\protect), which is the
general formula for the FCS of an interacting system and the main
result of this paper. We then proceed to evaluate the perturbative
corrections in $U$, see Section \ref{perttheory}, and investigate linear
response FCS on general grounds in Section \ref{linear} The opposite case
of large $U$, when the system is in the Kondo regime, is the
subject of Section \ref{KIFCS}. We not only present analytical
results at the Toulouse point, but also analyse the change in the
statistics around it in Section \ref{corraround}. Subsequently we
discuss the relation of the Kondo FCS to that of a RL set-up
between two LL at $g=1/2$ and establish connections to existing
results. Some conclusions are offered in Section
\ref{conclusions}. There are several appendices containing
technical details of some of the lengthier derivations.

\section{Keldysh method for the calculation of current statistics}
\label{SectionI}

\subsection{General considerations}

The cumulants of a given distribution function are known to define the
latter in the unique way \cite{papoulis}. For practical reasons it is usually
more convenient to calculate the so-called generating function
$\chi(\lambda)$, which in case of charge transport is
given by $\chi(\lambda) = \sum_q \, e^{i q \lambda} P_q$, where
$P_q$ is the probability for the charge $q$ to be
transferred through the system during the measuring time $\cal T$.
The parameter $\lambda$ here
is referred to as the measuring field.
The cumulants $\langle \delta^n q \rangle$
(which are nothing else but the irreducible moments of $P_q$)
can then be found for according to the prescription
\begin{eqnarray} \nonumber
 \langle \delta^n q \rangle = (-i)^n \left. \frac{\partial^n}{\partial
 \lambda^n} \,
 \ln \chi (\lambda)\right|_{\lambda = 0} \, .
\end{eqnarray}
The measurement of the charge transmitted through a system is
usually accomplished by a coupling to a `measuring device'. In the
original work by Levitov and Lesovik it is a fictitious spin-$1/2$
galvanometer coupled to the current \cite{levitovlesovik,lll}. The
transmitted charge is then proportional to the net change of the
spin phase. As has been shown by Nazarov \cite{nazarovlong}, the
counting of charge can in general be done by coupling the system
to a fictitious field and calculating the non-linear response,
which leads, of course, to exactly the same results.

According to \cite{reznikov} the generating function is given by the
following average,
\begin{equation} \label{chi}
\chi(\lambda)= \left\langle T_{{\rm C}} \exp
\left[-i\int\limits_{{\rm C}} T_\lambda(t)dt \right]
\right\rangle\;,
\end{equation}
where ${\rm C}$ is the Keldysh contour, $T_C$ is the contour
ordering operator, $\lambda(t)$ is the {\it measuring field} which
is non-zero only during the measuring time ${\cal
  T}$:
$\lambda(t)=\lambda\theta (t)\theta({\cal T}-t)$ on the forward path
and $\lambda(t)=-\lambda\theta (t)\theta({\cal T}-t)$ on
the backward path.
Introducing the operator transferring an electron through the system in the
positive direction (i.~e. in the direction of the current) $T_R$, and its
counterpart $T_L$ we can write
\begin{eqnarray}  \label{Tlambdaoperator}
T_\lambda= e^{i\lambda(t)/2}T_R+e^{-i\lambda(t)/2}T_L\;.
\end{eqnarray}
We note in passing that $T_R^\dagger=T_L$
in any system. Consequently, writing out (\ref{chi}) explicitly in
terms of the time--ordered and anti--time--ordered products, one arrives
at the conjugation property
\begin{equation} \label{conj}
\chi^*(\lambda)=\chi(-\lambda)\;.
\end{equation}
We now allow $\lambda(t)$ to be an arbitrary function on the
Keldysh contour, $\lambda_\mp(t)$ on the forward/backward path.
Then a generalised counterpart of (\ref{chi}) can be defined as
\begin{equation} \label{chipm}
  \chi[\lambda_-(t),\lambda_+(t)]=\langle T_{\rm C}
  e^{-i\int_{{\rm C}} \, dt \,
  T_\lambda(t)}\rangle \, .
\end{equation}
Next we assume that the measuring field changes only very slowly in time.
Then up to
the switching terms (which are known to be proportional to $\ln {\cal T}$)
\begin{equation} \label{potential} \nonumber
\chi[\lambda_-(t),\lambda_+(t)]=\exp\left[-i\int_0^{\cal T} {\cal
U}[\lambda_-(t),\lambda_+(t)]dt \right]
\end{equation}
where ${\cal U}(\lambda_-,\lambda_+)$ is the {\it adiabatic
potential}. Once the adiabatic potential is computed, the
statistics is recovered from
\begin{equation} \nonumber
\ln \chi(\lambda)=-i {\cal T}\, {\cal U}(\lambda,-\lambda)\;.
\end{equation}
Alternatively we can level off the $\lambda_\pm$ functions in
Eq.~(\ref{chipm}) to different constants as
\begin{equation} \nonumber
\chi[\lambda_-(t),\lambda_+(t)]\to\chi(\lambda_-,\lambda_+)\;,
\end{equation}
then $\chi(\lambda)=\chi(\lambda,-\lambda)$. Note that the
conjugation property (\ref{conj}) now generalises to
\begin{equation}\label{conj2} \nonumber
\chi^*(\lambda_-,\lambda_+)=\chi(\lambda_+,\lambda_-)\;,
\end{equation}
or
\begin{equation} \nonumber
{\cal U}^*(\lambda_-,\lambda_+)=-{\cal U}(\lambda_+,\lambda_-) \, .
\end{equation}

To calculate the adiabatic potential we observe that according to
the non-equilibrium version of the Feynman--Hellmann theorem
\cite{FH,ng,ourPhonon},
\begin{equation}\label{FHel}
\frac{\partial}{\partial\lambda_-}{\cal U}(\lambda_-,\lambda_+)=
\left\langle \frac{\partial T_\lambda(t)}{\partial\lambda_-}
\right\rangle_\lambda\;,
\end{equation}
where we use notation
\begin{equation} \nonumber
\langle A(t) \rangle_\lambda=\frac{1}{\chi(\lambda_-,\lambda_+)}
\left\langle T_{{\rm C}}\left\{A(t)e^{-i\int\limits_{{\rm C}}
T_\lambda(t)dt} \right\} \right\rangle
\end{equation}
(and similarly for multi--point averages) where $\lambda$'s are
understood to be {\it different} constants on the two time
branches. Note that the above one--point averages depend on the
branch the time $t$ is on (though not on the value of $t$ on that branch):
\begin{equation} \nonumber
\langle A(t_-) \rangle_\lambda\neq \langle A(t_+) \rangle_\lambda \, .
\end{equation}
Therefore the average in Eq.~(\ref{FHel}) {\it must} be taken on
the forward branch of the Keldysh contour. One immediate advantage
of our Hamiltonian approach is the fact that the calculation of
the adiabatic potential ${\cal U}$ amounts to a calculation of
some well defined Green's function (GF), even though a
non-equilibrium one. So we can use the whole power of the diagram
technique and connect to many known results within this method
without being restricted to scattering formalism as in
\cite{lll,reznikov}.

\subsection{FCS of a tunnelling junction}
\label{tunnjunction}

In order to illustrate the procedure we calculate the FCS of the
tunnelling junction between two metallic electrodes, denoted by
$R$ and $L$, which we model by the wide flat band Hamiltonians
$H_0[\psi_{R,L}]$. Their chemical potentials are assumed to be
$\mu_{R,L} = \pm V/2$, where $V$ is the voltage applied across the
junction (we set $e=m=\hbar=k_B=1$ and the Fermi energy $E_F=0$
throughout). The coupling between the electrodes is supposed to be
the conventional point-like tunnelling with the amplitude
$\gamma$, so that (for simplicity we assume spin-less electrons)
\begin{eqnarray} \nonumber
 H = \sum_{i=R,L} H_0[\psi_i] + \gamma \left[ \psi_R^\dag(0) \psi_L(0) +
 \mbox{h.c.}\right] \, .
\end{eqnarray}
The unperturbed GFs (for $\gamma=0$) can be easily evaluated, see
e.~g. \cite{mahan} ($i=R,L$),
\begin{eqnarray}   \label{bareGFs}
  g_i^{--}(\omega) &=& g_i^{++}(\omega)=  i 2 \pi \rho_0
 [n_i-1/2] \, , \nonumber \\
 g_i^{-+}(\omega) &=& i 2 \pi \rho_0 n_i \; ,
 \nonumber \\
 g_i^{+-}(\omega)
 &=& - i 2 \pi \rho_0 [1 -  n_i] \; ,
\end{eqnarray}
where $\rho_0$ is the density of states in the electrodes in
the vicinity of $E_F$. Here $n_{R,L}=n_F(\omega \pm V/2)$ where $n_F$ is
the Fermi distribution function. We use the original notation of
Keldysh for the GFs, where the superscripts stand for the position
of the time arguments on the contour $C$ rather than the far more
widespread language in terms of retarded (advanced) and
thermodynamic components (\cite{keldysh,LLX} vs.
\cite{rammersmith,langreth}). The reason for that is the fact that
due to the presence of two different fields $\lambda_\pm$ the
fundamental relation connecting the four Keldysh GFs, $G^{--} +
G^{++} = G^{-+} + G^{+-}$, does not hold any more. Therefore in
the present situation there are indeed four {\it independent} GFs.

For obvious reasons the $T_\lambda$ operator (\ref{Tlambdaoperator}) is given
by \cite{reznikov}
\begin{equation} \nonumber
 T_\lambda = \gamma \left[ e^{i \lambda} \psi_R^\dag(0) \psi_L(0) +
   e^{-i \lambda} \psi_L^\dag(0) \psi_R(0)\right] \, ,
\end{equation}
so that we have to evaluate
\begin{equation}       \label{longU}
 \frac{\partial}{\partial\lambda_-}{\cal U}(\lambda_-,\lambda_+)=
i\gamma \langle
e^{i\lambda_-}\psi_R^\dagger \psi_L - e^{-i\lambda_-}\psi_L^\dag \psi_R
\rangle_\lambda\;.
\end{equation}
Defining the mixed GFs,
\begin{eqnarray}
 G_{RL}(t,t') &=& - i \langle T_C \psi_R(t) \psi_L^\dag(t') \rangle_\lambda \,
 ,  \nonumber \\ \nonumber
 G_{LR}(t,t') &=&
- i \langle T_C \psi_L(t) \psi_R^\dag(t') \rangle_\lambda \, ,
\end{eqnarray}
we can re-write (\ref{longU}) as
\begin{eqnarray}                       \label{Uone}
  \frac{\partial}{\partial\lambda_-}{\cal U}(\lambda_-,\lambda_+)=
 \lim_{\epsilon \rightarrow 0^+} \int \frac{d \omega}{2 \pi} e^{i
  \epsilon \omega} \nonumber \\
 \times
 \left[ \gamma^* e^{i \lambda_-} G_{LR}^{--}(\omega) - \gamma e^{-i \lambda_-}
  G_{RL}^{--}(\omega) \right] \, .
\end{eqnarray}
The calculation of the GFs $G_{LR(RL)}^{--}(\omega)$ is most
elegantly accomplished using functional integration. To that end
we introduce the matrix of GFs according to
\begin{eqnarray}          \label{4x4}    \nonumber
\hat{g}=\left[
\begin{array}{cccc}
g_{RR}^{--} & g_{RR}^{-+} & g_{RL}^{--} & g_{RL}^{-+} \\
g_{RR}^{+-} & g_{RR}^{++} & g_{RL}^{+-} & g_{RL}^{++} \\
g_{LR}^{--} & g_{LR}^{-+} & g_{LL}^{--} & g_{LL}^{-+} \\
g_{LR}^{+-} & g_{LR}^{++} & g_{LL}^{+-} & g_{LL}^{++} \\
\end{array}\right] \;.
\end{eqnarray}
Using (\ref{bareGFs}) one easily constructs the corresponding
matrix $\hat{g}_0$ without tunnelling. The GFs for $\gamma \neq 0$
are found from the equation
\begin{eqnarray}               \label{eqnO}
\hat{G}^{-1} = \hat{g}_0^{-1} - _T\hat{\Sigma} \, ,
\end{eqnarray}
where $_T\hat{\Sigma}$ is the self-energy due to tunnelling. It
is found to have only four non-zero components,
\begin{eqnarray} \nonumber
\begin{array}{ll}
_T\hat{\Sigma}^{--}_{RL} = \gamma e^{i \lambda_-} &
_T\hat{\Sigma}^{++}_{RL} = - \gamma e^{i \lambda_+} \\
_T\hat{\Sigma}^{--}_{LR} = \gamma^* e^{-i \lambda_-} &
_T\hat{\Sigma}^{++}_{LR} = -\gamma^* e^{-i \lambda_+}
\end{array} \, ,
\end{eqnarray}
Solving (\ref{eqnO}) results in
\begin{eqnarray}
 \mbox{Det} \, (\hat{g}_0^{-1} &-& _T\hat{\Sigma}) G_{RL}^{--}
\nonumber \\ \nonumber
 &=&  - e^{i \lambda_-}
 \frac{\gamma}{(\pi \rho_0)^{2}}
 [1 + \Gamma + 2 (n_R - n_L)] \, , \nonumber \\
 \mbox{Det} \, (\hat{g}_0^{-1} &-& _T\hat{\Sigma}) G_{LR}^{--} =
 - e^{-i \lambda_-} \frac{\gamma^*}{(\pi \rho_0)^{2}} \nonumber \\
 \nonumber
&\times& [1+\Gamma + 2 (1-2e^{i\bar{\lambda}})(n_R-n_L)] \, ,
\end{eqnarray}
where $\bar\lambda = \lambda_- - \lambda_+$ and $\Gamma=(\pi
\rho_0 \gamma)^2$ is the dimensionless contact transparency. The
determinant is found to be given by
\begin{eqnarray} \nonumber
&& \mbox{Det} \, (\hat{g}_0^{-1} - _T\hat{\Sigma}) = (\pi
\rho_0)^{-4}\Big\{ (1 + \Gamma)^2  + 4 \Gamma  \\ \nonumber
&\times& \Big[ (e^{i \bar\lambda}- 1 ) n_L(1 - n_R) + (e^{-i
\bar\lambda}- 1 ) n_R(1 - n_L)\Big] \Big\} \, .
\end{eqnarray}
Inserting this outcome into (\ref{Uone}), integrating over
$\lambda_-$ and setting $\lambda_- = - \lambda_+ = \lambda$ we
immediately arrive at the Levitov--Lesovik formula
\begin{eqnarray}                            \label{LLformula}
 &~&\ln \chi_{0}(\lambda;V;\{T(\omega)\}) =  {\cal T} \,
 \int  \frac{d \omega}{2 \pi}
 \ln \Big\{
1 + T(\omega)\\ &\times&\big[ n_L(1-n_R)
(e^{i \lambda}-1) +
 n_R(1-n_L)(e^{-i \lambda}-1) \big] \Big\} \,,\nonumber
\end{eqnarray}
where the transmission coefficient is given by $T(\omega) = 4
\Gamma/(1+\Gamma)^2$ for the particular case of the tunnelling
junction set-up. Eq.~(\ref{LLformula}) holds of course, for any
non-interacting system, with known transmission coefficient,
coupled to two non-interacting reservoirs described by filling
factors $n_{R,L}$.

The generating function (\ref{LLformula}) leads at zero
temperature and small voltage to the conventional binomial
distribution function
\begin{eqnarray}      \label{binomialdistribchi} \nonumber
 \chi(\lambda) = \left[ 1-T(0) + T(0)e^{i \lambda} \right]^N \, ,
\end{eqnarray}
where $N={\cal T} V/2 \pi = {\cal T} V e^2/h$ is the number of
incoming particles during the waiting time (also known as `number
of attempts') and $1-T(0)$ and $T(0)$ are their probabilities to
be reflected or transmitted, respectively. Generally, the terms
proportional to $(e^{i m \lambda}-1)$ may be interpreted as
describing the tunnelling processes of particles with the
elementary charge $me$ \cite{levitovlesovik}. Negative $m$
correspond then to transport in direction opposite to that of the
applied voltage. Due to the detailed balance principle, such terms
do not contribute at $T=0$.

\section{FCS of the Anderson impurity problem}
\label{FCSAI}

\subsection{Preliminaries}

Now we are in a position to proceed to more complicated models.
The Hamiltonian of the AIM model consists of three
contributions,
\begin{eqnarray}\label{hamand} \nonumber
 H = H_0 + H_T + H_C \, .
\end{eqnarray}
The kinetic part
\begin{equation}                \label{kinpart}  \nonumber
 H_0 = \sum_\sigma H_0[\psi_{R/L,\sigma}]+\sum_\sigma(\Delta_0+
\sigma h)d^\dagger_\sigma d_\sigma \, ,
\end{equation}
describes a single fermionic level (which we shall also call `dot')
with electron creation operators
$d^\dag_\sigma$ ($\sigma$ is the spin index), energy $\Delta_0$ and subject
to a local magnetic field $h$. Two non-interacting
metallic leads $i=R,L$ are modelled as in the previous Section.
The leads and the dot are coupled via tunnelling,
\begin{equation}                 \label{H_T} \nonumber
 H_T = \sum_\sigma\left[
\gamma_L e^{i\lambda(t)/2}d^\dagger_\sigma \psi_{L\sigma}
+\gamma_R \psi^\dagger_{R\sigma}
d_\sigma\psi^\dagger_{R\sigma}+{\rm H.c.}\right] \, ,
\end{equation}
with different amplitudes $\gamma_{R,L}$. For convenience we
already included the counting field into the Hamiltonian. Notice
that since the transfer of a physical electron through the device
is a two-stage process (left lead $\rightarrow$ dot $\rightarrow$
right lead or the other way round) the measuring field is halved.
For the sake of simplicity we incorporate the counting field only
into the left junction. Doing that at both junctions (of course
with the correction $\lambda/2 \rightarrow \lambda/4$) leads to
exactly the same results due to the gauge symmetry of the
Hamiltonian. Finally, we include the Coulomb repulsion on the dot,
\begin{equation}                 \label{H_C} \nonumber
 H_C = Un_\uparrow n_\downarrow\; ,
\end{equation}
where $n_\sigma=d^\dagger_\sigma d_\sigma$. The applied voltage is
incorporated into the full Hamiltonian as in the previous Section,
$\mu_L-\mu_R=V\geq 0$.

We start with the definition of two auxiliary GFs,
\begin{eqnarray}
 F_\lambda(t,t')
&=& -i\langle T_{{\rm C}}\{\psi_L(t)d^\dagger(t')\}\rangle_\lambda\; \nonumber
 \\ \nonumber \widetilde{F}_\lambda(t,t')
&=& -i\langle T_{{\rm
C}}\{d(t)\psi^\dagger_L(t')\}\rangle_\lambda\; \,
\end{eqnarray}
Hence the derivative of the adiabatic potential is given by
\begin{eqnarray}                      \label{UF}
\frac{\partial}{\partial\lambda_-}{\cal U}(\lambda_-,\lambda_+)=
\frac{\gamma_L}{2}\lim\limits_{\epsilon\to 0+}
\int\frac{d\omega}{2\pi}e^{i\epsilon\omega}
\nonumber \\ \times
\left[e^{i\lambda_-/2}F^{--}_\lambda(\omega)-e^{-i\lambda_-/2}
\widetilde{F}^{--}_\lambda(\omega)\right]\;.
\end{eqnarray}
Similar to the situation of the tunnelling junction these mixed
GFs can be written as combinations of bare lead GFs and exact
impurity GF $D(t,t)$,
\begin{eqnarray}
 \widetilde{F}_\lambda(t,t') &=& \int\limits_{{\rm C}}dt''g_L(t-t'')
 e^{-i \lambda (t'')} D(t'',t') \, , \nonumber \\ \nonumber
F_\lambda(t,t') &=& \int\limits_{{\rm C}}dt''D(t,t')
e^{-i \lambda (t'')} g_L(t''-t')\;.
\end{eqnarray}
Performing the Keldysh disentanglement and plugging the result back into
(\ref{UF}) one obtains
\begin{eqnarray}                   \label{ULR}
&& \frac{\partial}{\partial\lambda_-}{\cal U}(\lambda_-,\lambda_+)
\\
&=& \frac{\gamma_L^2}{2} \int \frac{d\omega}{2\pi} \Big[e^{-i
\bar\lambda/2}D^{-+} g^{+-}_L \nonumber \ -e^{i
\bar\lambda/2}g^{-+}_L D^{+-} \Big] \, ,
\end{eqnarray}
where again $\bar\lambda = \lambda_- - \lambda_+$. Hence, the
whole problem is now reduced to calculation of the impurity GF.
The most compact way to access it is using the self-energy
formalism. According to \cite{LLX} the self-energy
$\hat{\Sigma}(\omega)$ for a non-equilibrium system can be defined
in very much the same way as in the traditional diagram technique
via
\begin{equation}     \label{selfenergyequation}
 \hat{D}(\omega) = \hat{d}_0(\omega) + \hat{D}(\omega) \hat{\Sigma} (\omega)
 \hat{d}_0 (\omega) \, ,
\end{equation}
where the unperturbed dot GF is
\begin{eqnarray}                \label{cleardot}
  \hat{d}_0^{-1}=\left[\begin{array}{lc}\omega-\Delta_0 & 0 \\
  0 &-\omega+\Delta_0\end{array}\right]\;.
\end{eqnarray}
Then trivially
\begin{equation} \nonumber
 \hat{D}^{-1}(\omega) = \hat{d}^{-1}_0 (\omega) -  \hat{\Sigma} (\omega) \, .
\end{equation}
Therefore our goal now is the evaluation of the self-energy.

\subsection{The $U=0$ case: resonant level model}
\label{RLFCS}

We shall elaborate on the formula (\ref{ULR}), which is still
valid for the interacting case, in the following subsection. For
pedagogical reasons we pause here to deal with $U=0$ case, when
$H$ {\it is trivially diagonalisable}. This situation is referred
to as the resonant level (RL) model.

The corresponding self-energy is (we neglect the spin index here
as GFs are diagonal in $\sigma$ and independent of it, the
sub-script `0' distinguishes the $U=0$ quantities):
\begin{widetext}
\begin{eqnarray}
\hat{\Sigma}_0(\omega)&=&
\left[
\begin{array}{lr}
\gamma_L^2 g^{--}_L+\gamma^2_R g_R^{--} & -e^{i
\bar\lambda/2}\gamma_L^2 g_L^{-+}-\gamma_R^2 g^{-+}_R \\ -e^{-i
\bar\lambda /2}\gamma_L^2 g_L^{+-}-\gamma_R^2 g^{+-}_R& \gamma_L^2
g^{++}_L+\gamma^2 g_R^{++}
\end{array}
\right]\nonumber \nonumber \\ &=&\left[\begin{array}{lr} i\Gamma_L
(2n_L-1)+i\Gamma_R (2n_R-1) & -2ie^{i \bar\lambda/2}\Gamma_L
n_L-2i\Gamma_R n_R \\ 2ie^{-i\bar\lambda/2}\Gamma_L
(1-n_L)+2i\Gamma_R (1-n_R)& i\Gamma_L (2n_L-1)+i\Gamma_R (2n_R-1)
\end{array}
\right]\;,
\nonumber
\end{eqnarray}
where, in order to unburden the notation, we set
$\Gamma_{R,L}= (\pi \rho_0 \gamma_{R,L})^2$. Consequently
\begin{eqnarray}                    \label{simpleD}
\hat{D}_0^{-1}(\omega)= \left[\begin{array}{lr} \omega-\Delta_0
-i\Gamma_L (2n_L-1)-i\Gamma_R (2n_R-1) & 2ie^{i
\bar\lambda/2}\Gamma_L n_L+2i\Gamma_R n_R \\ -2ie^{-i
\bar\lambda/2}\Gamma_L (1-n_L)-2i\Gamma_R (1-n_R)&
-\omega+\Delta_0-i\Gamma_L (2n_L-1)-i\Gamma_R (2n_R-1)
\end{array}
\right]\;.
\end{eqnarray}
Inversion of this results in
\begin{eqnarray}
\hat{D}_0(\omega)&=&\frac{1}{{\cal D}_0(\omega)}\label{DbareU}\\
&\times&  \left[ \begin{array}{lr} \omega-\Delta_0 +i\Gamma_L
(2n_L-1)+i\Gamma_R (2n_R-1) & 2ie^{i \bar\lambda/2}\Gamma_L
n_L+2i\Gamma_R n_R \\ -2ie^{-i \bar\lambda/2}\Gamma_L
(1-n_L)-2i\Gamma_R (1-n_R)& -\omega+\Delta_0+i\Gamma_L
(2n_L-1)+i\Gamma_R (2n_R-1)
\end{array}
\right]\;,\nonumber
\end{eqnarray}
where ($\Gamma = \Gamma_R + \Gamma_L$)
\begin{equation}\label{denominatorU=0}
{\cal D}_0(\omega)=(\omega-\Delta_0)^2+\Gamma^2+4\Gamma_L\Gamma_R
\left[n_L(1-n_R)(e^{i \bar\lambda/2}-1)+n_R(1-n_L)(e^{-i
\bar\lambda/2}-1)\right]\;.
\end{equation}
Inserting these results back into (\ref{ULR}) yields an equation,
\begin{eqnarray}                \label{reslevelbeforeintegration}
\frac{\partial}{\partial\lambda_-}{\cal U}(\lambda_-,\lambda_+)=
-2\Gamma_L\Gamma_R
\int\limits_{-\infty}^{\infty}\frac{d\omega}{2\pi}
\frac{n_L(1-n_R)e^{i \bar\lambda/2} - n_R(1-n_L)e^{-i
\bar\lambda/2} }{{\cal D}_0(\omega)}\; \;.
\end{eqnarray}
Performing the integration over $\lambda_-$ and constructing the generating
function we again find the formula (\ref{LLformula}) with the Breit-Wigner
transmission coefficient
\begin{eqnarray} \nonumber
 T(\omega)=\frac{4\Gamma_L\Gamma_R}{(\omega-\Delta_0)^2+\Gamma^2} \, ,
\end{eqnarray}
as expected for the RL set-ups \footnote{The related problem of a
double-barrier junction has been analysed in \cite{dejong}.}.

\subsection{The general formula}
\label{general}

The GFs (\ref{DbareU}) can be used to construct the consistent
expansion of the FCS to {\it all powers} in $U$, opening the road
to perturbative as well as non-perturbative studies of the FCS.
From now on under $\hat{\Sigma}$ we shall understand the
self-energy due to the Coulomb interaction (tunnelling terms are
incorporated into the bare GFs). Eq.~(\ref{simpleD}) thus changes
to
\begin{eqnarray}                        \label{D0beforeinversion}
 \hat{D}^{-1}(\omega)=
\left[\begin{array}{lr} \omega-\Delta_0 -i\Gamma_L
(2n_L-1)-i\Gamma_R (2n_R-1) -\Sigma^{--} & 2ie^{i
\bar\lambda/2}\Gamma_L n_L+2i\Gamma_R n_R -\Sigma^{-+} \\ -2ie^{-i
\bar\lambda/2}\Gamma_L (1-n_L)-2i\Gamma_R (1-n_R) -\Sigma^{+-}&
-\omega+\Delta_0-i\Gamma_L (2n_L-1)-i\Gamma_R (2n_R-1)-
\Sigma^{++}
\end{array}
\right]\;
\end{eqnarray}
After the inversion of this matrix and insertion it into
(\ref{ULR}) one gets [${\cal D}(\omega)$ is the corresponding
counterpart to (\ref{denominatorU=0})]
\begin{eqnarray}             \label{fi}
\frac{\partial}{\partial\lambda_-}{\cal U}(\lambda_-,\lambda_+)&=&
-\Gamma_L\int\limits_{-\infty}^\infty\frac{d\omega}{2\pi{\cal
D}(\omega)} \left\{2\Gamma_R\left[e^{i \bar\lambda/2}
n_L(1-n_R)-e^{-i \bar\lambda/2} n_R(1-n_L)\right]
 \right. \\
&-& i \left. \left[ e^{i \bar\lambda/2}n_L\Sigma^{+-}+e^{-i
\bar\lambda/2} (1-n_L)\Sigma^{-+}\right]\right\}\;,\nonumber
\end{eqnarray}
which is a general formula for the statistics in interacting
systems. Here ${\cal D}(\omega)$ is the, $\lambda$-dependent,
determinant of the matrix given by Eq.~(\ref{D0beforeinversion}).
For $\bar\lambda=0$ the rhs of this relation is proportional to
the current through the device. Moreover, as expected, in this
particular case Eq.~(\ref{fi}) can be brought into the form
derived by Meir--Wingreen \cite{meirwingreen}, when the transport
is defined solely by the retarded dot level GF after a
symmetrisation procedure. The presence of the counting field does
not allow a similar reduction for arbitrary $\lambda$ though.

Clearly formula (\ref{fi}) is not restricted to the AIM as such
but is applicable for any similar one-channel impurity set-up
(including, e.~g. electron--phonon interaction on the dot or a
double dot).

\subsection{Perturbative expansion in the Coulomb interaction}
\label{perttheory}

The obvious way to proceed is to calculate the lowest-order
contributions to the self-energy, in the time domain
\begin{equation}\label{sigma12}
\hat{\Sigma}(t)=\left[
\begin{array}{lr}
-iUD_0^{--}(0) + U^2 [D_0^{--}(t)]^2D_0^{--}(-t) & -U^2
[D_0^{-+}(t)]^2D_0^{+-}(-t) \nonumber \\ \nonumber -U^2
[D_0^{+-}(t)]^2D_0^{-+}(-t) & iUD_0^{++}(0)+U^2
[D_0^{++}(t)]^2D_0^{++}(-t)
\end{array}
\right] \, .
\end{equation}
The linear in $U$ part is diagonal and is essentially a remnant of the
occupation probability of the dot level $\langle d^\dag d \rangle$.
It is most
conveniently evaluated in the following way (from now on we consider a
symmetrically coupled system $\Gamma_R=\Gamma_L=\Gamma/2$ at zero temperature
in order to simplify the algebra)
\begin{eqnarray} \nonumber
 -iUD_0^{--}(0) = - i U D_0^{-+}(0) = - i U \int \, \frac{d \omega}{2 \pi}
  D_0^{-+}(\omega) = U n_\lambda \, ,
\end{eqnarray}
where the object
\begin{eqnarray} \nonumber
 n_\lambda = \frac{1}{2 \pi} \left\{ (1+e^{i \lambda})
 \left[\frac{\pi}{2} -
 \tan^{-1} \left( \frac{\Delta_0 + V/2}{\Gamma} \right)
 \right] + e^{i \lambda/2}
 \sum_\pm \pm \tan^{-1} \left[ \frac{(\Delta_0 \pm V/2)e^{- i
 \lambda/2}}{\Gamma}
\right] \right\} \, ,
\end{eqnarray}
is, in general, $\lambda$ dependent. For $D_0^{-+}$ see
Eq.~(\ref{DbareU}). Here $n_0$ simply gives the dot occupation
probability. Plugging this result into (\ref{D0beforeinversion})
and proceeding to (\ref{fi}) we find the result identical to
(\ref{reslevelbeforeintegration}) up to the denominator
(\ref{denominatorU=0}) where the bare level energy $\Delta_0$ now
gets renormalised, $\Delta_0 \rightarrow \Delta_0 + U n_\lambda$.
Subsequent expansion in $U$ and integration over energy results in
a well controlled contribution which vanishes for the case of
the symmetric Anderson model $\Delta_0 = -U/2$, to which case
the following considerations are restricted.

We concentrate now on the correction at the second order in $U$.
One way to access the self-energies is through the evaluation
of the corresponding susceptibilities. We define them as [note the
sub-script `0', not to confuse with the generating function
$\chi(\lambda)$]
\begin{eqnarray}                 \label{suscdefinition} \nonumber
\hat{\chi}_0(\Omega)=i\int\limits_{-\infty}^\infty\frac{d\omega}{2\pi}
\left[
\begin{array}{lc}
D_0^{--}(\omega+\Omega)D_0^{--}(\omega) &
D_0^{-+}(\omega+\Omega)D_0^{+-}(\omega) \\
D_0^{+-}(\omega+\Omega)D_0^{-+}(\omega) &
D_0^{++}(\omega+\Omega)D_0^{++}(\omega)
\end{array}
\right]\;.
\end{eqnarray}
The respective self-energy can be extracted from
\begin{eqnarray} \nonumber
\hat{\Sigma}(\omega)=i\int\limits_{-\infty}^\infty\frac{d\Omega}{2\pi}
\left[
\begin{array}{lc}
D_0^{--}(\omega-\Omega)\chi_0^{--}(\Omega) &
D_0^{-+}(\omega-\Omega)\chi_0^{+-}(\Omega)\\
D_0^{+-}(\omega-\Omega)\chi_0^{-+}(\Omega) &
D_0^{++}(\omega-\Omega)\chi_0^{++}(\Omega)
\end{array}
\right]\;.
\end{eqnarray}
The equilibrium results have been originally presented in famous
series of papers by Yosida--Yamada \cite{yamada1,yamada2,yamada3}
(we set $T=0$ for simplicity),
\begin{equation}\label{sigmaeq} \nonumber
\hat{\Sigma}_{{\rm eq}}(\omega)=(1-\chi_{{\rm e}})\omega
\left[\begin{array}{lr}1&0\\0&-1\end{array}\right]-
\frac{i\chi_{{\rm o}}^2}{2\Gamma}\omega^2
\left[\begin{array}{ll}{\rm sign}(\omega)&
2\theta(-\omega)\\-2\theta(\omega)&
{\rm sign}(\omega)\end{array}\right]\;,
\end{equation}
where the {\it exact} even--odd susceptibilities possess
the following expansions in powers of $U$,
\begin{eqnarray} \nonumber
\chi_{{\rm e}}=1+\left(3-\frac{\pi^2}{4}\right)
\frac{U^2}{\pi^2\Gamma^2}+...\;,\;\;\;
\chi_{{\rm o}}=-\frac{U}{\pi\Gamma}\;.
\end{eqnarray}

For finite $V$ and at the second order in $U$,
there are three distinct energy regions
contributing to Eq.~(\ref{fi}): $-V/2<\omega<V/2$,
$V/2<\omega<3V/2$, and $-3V/2<\omega<-V/2$.
The low-energy expansion (not only small $U$
but small $V$ as well) in presence of $\lambda$ one finds in the
region $-V/2<\omega<V/2$:
\begin{eqnarray}\label{sigma2} \nonumber
\hat{\Sigma}(\omega)&=&(1-\chi_{{\rm e}})\omega
\left[\begin{array}{lr}1&0\\0&-1\end{array}\right]\\&-&
\frac{iU^2}{8\pi^2\Gamma^3}
\left[\begin{array}{cc}6\omega V&
e^{-i\lambda}\left(\frac{3V}{2}-\omega\right)^2+
3\left(\frac{V}{2}-\omega\right)^2
\\-e^{-2i\lambda}\left(\frac{3V}{2}+\omega\right)^2-
3e^{-i\lambda}\left(\frac{V}{2}+\omega\right)^2&
6\omega V\end{array}\right]\;.\nonumber
\end{eqnarray}
Needless to say, these relations are consistent with the non-equilibrium
calculation by Oguri \cite{oguri}. On the other hand,
for $\omega > V/2$ one obtains
\begin{eqnarray}      \nonumber
 \Sigma^{-+}(\omega)=&-& \frac{ie^{-i\lambda}U^2}{8\pi^2\Gamma^3}
\left(\frac{3V}{2}-\omega\right)^2\theta\left(\frac{3V}{2}-\omega\right)
\, ,\label{sigma2b}
\end{eqnarray}
while for $\omega < V/2$ the relation
\begin{eqnarray}   \nonumber
 \Sigma^{+-}(\omega)= \frac{ie^{-2i\lambda}U^2}{8\pi^2\Gamma^3}
\left(\frac{3V}{2}+\omega\right)^2\label{sigma2bb}
\theta\left(\frac{3V}{2}+\omega\right) \,
\end{eqnarray}
holds.
These self-energies, being incorporated into (\ref{fi}), yield the following
generating function for the FCS,
 \begin{equation}\label{corfull}   \nonumber
\ln \chi(\lambda)= \ln \chi_0(\lambda)+
\frac{{\cal T}U^2V^3}{24\pi^3\Gamma^4}(e^{-i\lambda}-1)+
\frac{{\cal T}U^2V^3}{12\pi^3\Gamma^4}(e^{-2i\lambda}-1)+O(U^4)\;,
\end{equation}
where
\begin{eqnarray}     \nonumber
\ln \chi_0(\lambda)={\cal T}\int\limits_{-V/2}^{V/2}\frac{d\omega}{2\pi}
\ln \left[1+\frac{\Gamma^2}{\chi_{{\rm e}}^2\omega^2+\Gamma^2}
(e^{i\lambda}-1)\right] \,
\end{eqnarray}
still contains $U$. Performing the expansion around the perfect transmission
(hence the sign change of $\lambda$ in the following formulas)
we see that in terms of susceptibilities
\begin{equation}\label{corchi}
\ln \chi(\lambda)=N\left\{i\lambda+\frac{V^2}{3\Gamma^2}
\left[\frac{\chi_{{\rm e}}^2+\chi_{{\rm o}}^2}{4}(e^{-i\lambda}-1)
+\frac{\chi_{{\rm o}}^2}{2}(e^{-2i\lambda}-1)\right]
\right\}\;,
\end{equation}
where $N={\cal T} V/ \pi$ is the number of incoming particles
during the measuring time slice. This is, of course, {\it only valid at
the order} $U^2$. We speculate that the general formula for the
full FCS could be written in terms of the equilibrium
susceptibilities $\chi_{o,e}$ only. One possibility is the
generating function of the form
\begin{equation}\label{corchibis}
\chi(\lambda)=N\ln\left[1+\left(1- \frac{\chi_{{\rm
e}}^2+3\chi_{{\rm o}}^2}{12\Gamma^2}V^2 \right)(e^{i\lambda}-1)
+\frac{\chi_{{\rm o}}^2}{6\Gamma^2}V^2(e^{-i\lambda}-1)
\right]\;,
\end{equation}
as this expression reproduces the expansion Eq.~(\ref{corchi}). We
stress again that so far we have only shown that
Eq.~(\ref{corchibis}) holds at the second order in $U$ and beyond
that it is a mere {\it hypothesis}.

It is tempting to interpret the appearance of the double
exponential terms as an indication of a coherent tunnelling of
electron pairs (caution: similar terms would also appear for the
non-interacting RL model due to the energy dependence of the
transmission coefficient). In the Toulouse limit calculation below
we find further evidence for such interpretation.

\end{widetext}

\subsection{Linear response FCS}
\label{linear}

Here we would like to take a closer look onto the general formula
(\ref{fi}) at zero temperature and vanishing applied voltage. In
order to arrive at correct results one has to bear in mind that
the limits $V \rightarrow 0$ and $\omega \rightarrow 0$ do not
commute in the presence of the counting field. Indeed, calculating the
Keldysh determinant in both limits we see that
\begin{equation}\label{noncom1}
\lim\limits_{\omega\to0}\lim\limits_{V\to0} {\cal
D}_0(\omega,V,\lambda)=\Delta_0^2+\Gamma^2\;,
\end{equation}
but
\begin{equation}\label{noncom2}
\lim\limits_{V\to0}\lim\limits_{\omega\to0} {\cal
D}_0(\omega,V,\lambda)=\Delta_0^2+\Gamma^2+4\Gamma_L\Gamma_R
(e^{i\lambda}-1)\;.
\end{equation}

In fact, it is the second scheme we have to implement analysing
the first term in Eq.~(\ref{fi}). This leads to a transmission
coefficient type contribution to the generating function.

On the contrary, in the second term in Eq.~(\ref{fi}), which is
produced by the self-energy, not even the integration over
$\omega$ is restricted to $[0,V]$. As a matter of fact, due to
Auger type effects \cite{gadzukplummer,ourFE} one expects that
there are contributions to the current (and FCS) at all energies.
This effect is itself proportional to the applied voltage though,
and results therefore in non-linear corrections to the FCS. Hence
the energy integration can be regarded to be restricted to $[0,V]$
even in the second term in Eq.~(\ref{fi}). Moreover, since the
self-energy does not have external lines and all the internal
frequencies have to be integrated over, the limits $V \rightarrow
0$ and $\omega \rightarrow 0$ in this case commute. That means
that for the evaluation of the self-energy to the lowest order in
$V$ one is allowed to use the equilibrium GFs, calculated in
presence of the counting field $\lambda$, i.~e. (\ref{DbareU})
with $n_R = n_L = n_F$ and with the corresponding Keldysh
denominator (\ref{noncom1}). Therefore all diagonal Keldysh GFs
are equal to those in the equilibrium and all off-diagonal ones
are simply proportional to the same diagrams as in equilibrium.
Since any given off-diagonal self-energy diagram describes an
inelastic process, it should vanish for $\omega \rightarrow 0$ and
we arrive at a conclusion that
\begin{eqnarray}   \nonumber
\lim\limits_{\omega\to 0} \hat{\Sigma}(\omega) =
\mbox{Re} \, \Sigma^R(0) \left[
 \begin{array}{cc}
  1 & 0 \\
  0 & -1
 \end{array} \right] \,
\end{eqnarray}
even at finite $\lambda$. Eq.~(\ref{fi}) thus leads to the
fundamental result
\begin{eqnarray}          \label{fundformula}
\ln\chi(\lambda)=N\ln\left\{1+\frac{\Gamma^2}{[{\rm
Re}\Sigma^{(R)}(0)]^2 +\Gamma^2}(e^{i\lambda}-1)\right\} \, ,
\end{eqnarray}
or to $\ln\chi(\lambda) = i \lambda N$ for the symmetric Anderson
impurity model. In case of the \emph{asymmetrically coupled}
impurity, $\Gamma_R \neq \Gamma_L$ the numerator of
(\ref{fundformula}) modifies to $\Gamma_R \Gamma_L$ while the
denominator contains $(\Gamma_R + \Gamma_L)/2$ instead of
$\Gamma$.

The result (\ref{fundformula}) allows simple generalisations to
asymmetric systems in a magnetic field $h$. According to
\cite{yamada1,yamada2,yamada3} the real part of the self-energy is
given by
\[
{\rm Re}\Sigma^{(R)}_\sigma(0)=\chi_{{\rm
c}}\kappa+\sigma\chi_{{\rm s}}h\;,
\]
where $\chi_{{\rm c/s}}$ are exact charge/spin susceptibilities
(combinations of even/odd) and $\kappa\sim\Delta_0+U/2$ is a
particle--hole symmetry breaking field. Consequently
\begin{eqnarray}\label{kondobin}
\ln\chi(\lambda)=\frac{N}{2}\ln\left\{
\left[1+\frac{\Gamma^2}{[\chi_{{\rm c}}\kappa+\chi_{{\rm s}}h]^2
+\Gamma^2}(e^{i\lambda}-1)\right] \nonumber \right. \\
\left. \times \left[1+\frac{\Gamma^2}{[\chi_{{\rm
c}}\kappa-\chi_{{\rm s}}h]^2
+\Gamma^2}(e^{i\lambda}-1)\right]\right\}\;.
\end{eqnarray}
The enormous advantage of this formula is the fact, that the
susceptibilities can be calculated \emph{exactly} for any system
parameters with the help of the Bethe-Ansatz results
\cite{kawakamiokiji,tsvelickwiegmann,vigmantsvelik}.

Let us stress that the result (\ref{fundformula}) is not limited
to the AIM but will hold for any similar model, hence the {\it
binomial theorem}. It is clear in hindsight that all the
non-elastic processes fall out in the $T=0$ linear response limit.
Still it is a remarkable result that {\it all} moments have a
simple expression in terms of a single number: the effective
transmission coefficient. The binomial distribution is universal.
[For a multi--channel system modifications will be required as is
obvious from looking at Eq.~(\ref{kondobin})].

\begin{widetext}
\subsection{The Kondo regime}
\label{KIFCS}

The way to proceed further is to consider the case of very deep
$\Delta_0$ and strong Coulomb repulsion. In this limiting case the
system is in the Kondo regime and the dot can in good
approximation be considered to be permanently populated by a
single electron. It has been shown in \cite{kaminski}, that the
conventional Schrieffer-Wolf transformation \cite{SWolf}, which
maps the Anderson impurity Hamiltonian onto that of the Kondo
problem, also works out of equilibrium. The result is the
two-channel Kondo Hamiltonian
\[
 H = H_0 + H_J + H_V + H_M \, ,
\]
where, with $\psi_{\alpha, \sigma}$ are the electron field
operators in the $\alpha=R,L$ electrodes,
\begin{eqnarray}\label{Hkondo}
 H_0 &=& i \sum_{\alpha=R,L} \sum_{\sigma=\uparrow,\downarrow}
 \int \, dx \,
 \psi^\dag_{\alpha \sigma}(x) \partial_x \psi_{\alpha \sigma}(x) \, \, \, \, \, ,
\, \, \, \, \,
 H_J = \sum_{\alpha, \beta = R,L} \sum_{\nu=x,y,z}
 J_\nu^{\alpha \beta}
 s^\nu_{\alpha \beta} \tau^\nu \, , \nonumber \\
 H_V &=& (V/2) \sum_\sigma \int \, dx \, ( \psi^\dag_{L \sigma}
 \psi_{L \sigma}
 - \psi^\dag_{R \sigma} \psi_{R \sigma}) \, \, \, \, \,  ,
 \, \, \, \, \,
 H_M =
 -\mu_B g_i h \tau^z =
 - \Delta \tau^z \, .
\end{eqnarray}
$\mu_B$ is the Bohr's magneton, $g_i$ the gyromagnetic ratio and $h$ denotes
the local magnetic field, which is applied to the impurity spin.
Here $\tau^{\nu=x,y,z}$ are the Pauli matrices for the
impurity spin and
\[
s^\nu_{\alpha \beta}=\sum_{\sigma,\sigma'}\, \psi_{\alpha \sigma}^\dag(0) \,
\sigma^\nu_{\sigma \sigma'} \, \psi_{\beta \sigma'}(0) \, ,
\]
are the components of the electron
spin densities in (or across) the leads, biased by a finite voltage $V$.
The last term in Eq.~(\ref{Hkondo}) stands for the magnetic
field, $\Delta=\mu_B g_i h$.  We follow \cite{SH} and
assume $J_x^{\alpha \beta} =
J_y^{\alpha \beta} = J_\perp^{\alpha \beta}$,
$J_{z \pm} = (J_z^{LL} \pm J_z^{RR})/2$ and $J_z^{LR}=J_z^{RL}=0$.
The only transport process then allowed is the
spin-flip tunnelling (sometimes also called `exchange co-tunnelling'),
so that we obtain for the $T_\lambda$ operator
\begin{eqnarray}
T_\lambda = \frac{J_\perp^{RL}}{2}\left( \tau^+
e^{i\lambda(t)/2} \psi_{R
  \downarrow}^\dag \psi_{L \uparrow} + \tau^-
  e^{i \lambda(t) /2} \psi_{R
  \uparrow}^\dag \psi_{L \downarrow}
 + \tau^+ e^{-i \lambda(t)/2} \psi_{L
  \downarrow}^+ \psi_{R \uparrow} + \tau^- e^{-i \lambda(t)/2} \psi_{L
  \uparrow}^\dag \psi_{R \downarrow} \right) \, .\nonumber
\end{eqnarray}
Of course, there is also a regular elastic co-tunnelling term, which couples
the leads directly. However, it can be rigorously shown \cite{kaminski},
that these processes are subleading in the low energy sector in comparison
to spin-flip tunnelling. That is why we keep only the latter contributions to
the Hamiltonian. We proceed by bosonization,
Emery-Kivelson rotation, and refermionization \cite{EK,SH,book}.
We obtain then with
$J_\pm = (J_\perp^{LL} \pm J_\perp^{RR})/\sqrt{2\pi a_0}$,
$J_\perp = J_\perp^{RL}/\sqrt{2 \pi a_0}$ ($a_0$ is the lattice constant of
the underlying lattice model)
\begin{eqnarray}                       \label{KondoHam}
  H &=&  i \sum_{\nu=c,s,cf,sf}
 \int \, dx \,
 \psi^\dag_{\nu}(x) \partial_x \psi_{\nu}(x) \,
 + \frac{J_+}{2} \left[ \psi_{sf}^\dag(0) +  \psi_{sf}(0)\right] \, \tau^y
 \nonumber \\
 &+&
 \left\{
 \frac{J_\perp}{2} \left[ \psi_{cf}^\dag(0) - \psi_{cf}(0)\right]
 +
 \frac{J_-}{2} \left[ \psi_{sf}^\dag(0) - \psi_{sf}(0)\right] \right\}
\, \tau^x
 \nonumber \\
 &-& \left[ (J_{z+} - 2 \pi): \psi_s^\dag(0) \psi_s(0): +
 J_{z-}: \psi_{sf}^\dag(0) \psi_{sf}(0): \right] \, \tau^z
 - \Delta \tau^z   + V \int \, dx \, \psi_{cf}^\dag(x) \psi_{cf}(x) \, ,
\end{eqnarray}
where now four fermionic channels are present: ($c$) total charge density
channel for the sum of particle densities in both electrodes,
($cf$) charge flavour channel for the
difference in densities. The channel-symmetric spin density channel ($s$)
and channel-antisymmetric (or spin flavour channel) ($sf$)
(see details in \cite{SH} and \cite{book}) are defined in analogy to their
charge counterparts.
A considerable simplification of the theory is
achieved by introduction of the Majorana components of the continuum fields
\begin{eqnarray}                    \label{Majoranashow}
 \eta_\nu =(\psi_\nu^\dag + \psi_\nu)/\sqrt{2} \; , \;
 \xi_\nu = i(\psi_\nu^\dag - \psi_\nu)/\sqrt{2} \, ,
\end{eqnarray}
and of the impurity spin $\tau^x = b$ and $\tau^y = a$. As a result,
the model
simplifies to (for convenience we kept $\psi_{s,sf}$ operators in the terms
quartic in fermions)
\begin{eqnarray}                         \label{2chmajham}
 H &=& H' + H'' \\ \nonumber
 H' &=& H'_0 - i (J_- \, b \, \xi_{sf} + J_+ \, a \, \eta_{sf}) - i
 \Delta \, a \, b + T_\lambda \\ \nonumber
 H'' &=& i \left[ v_1 :\psi_s^\dag(0) \psi_s(0): + J_{z-} :\psi_{sf}^\dag(0)
\psi_{sf}(0): \right] a \, b
\;,
\end{eqnarray}
where $v_1=J_{z+}-2 \pi$ and the counting term is given by
\begin{eqnarray}                \label{TlambdaKondo}
T_\lambda =
- i J_\perp b \, \left[ \xi_{cf} \, \cos (\lambda/2) - \eta_{cf} \, \sin
(\lambda/2) \right] \, .
\end{eqnarray}
The fields $\eta_{sf}$ and $\xi_{sf}$ in the spin--flavour sector
are equilibrium Majorana fields, whereas $\eta_{cf}$ and
$\xi_{cf}$ in the charge--flavour sector are biased by $V$ (from
now on we omit the $cf$ index and denote $sf$ by $f$),
\begin{eqnarray}                      \label{H0prime}   \nonumber
 H_0' = i \int\, dx \, \Big[ \eta_{f}(x)
 \partial_x \eta_{f}(x) + \xi_{f}(x) \partial_x \xi_{f}(x)
 + \eta_{}(x) \partial_x \eta_{}(x) + \xi_{}(x) \partial_x \xi_{}(x)
    + V \xi_{}(x) \eta_{}(x) \Big] \, ,
\end{eqnarray}
where we drop the $c$ and $s$ channels as they decouple from the
impurity completely (at the Toulouse point). The evaluation of the
adiabatic potential can now be performed along the lines of
Section \ref{SectionI},
\begin{eqnarray}                \label{firstU}
\frac{\partial}{\partial\lambda_-}{\cal U}(\lambda_-,\lambda_+)=
-\frac{J_\perp}{2}
\int \frac{d\omega}{2\pi}
\left[\sin(\lambda_-/2) \, G_{b\xi}^{--}(\omega)+ \cos(\lambda_-/2)
\, G_{b\eta}^{--}(\omega)\right]\;,
\end{eqnarray}
where we again define the mixed GFs according to the prescription
\begin{eqnarray}                      \label{KondoGFs}
 G_{b \xi}(t,t') = - i \langle T_C b(t) \xi(t) \rangle_\lambda \, \, \, \, \, ,
 \, \, \, \, \, G_{b \eta}(t,t') = - i \langle T_C b(t) \eta(t)
\rangle_\lambda \, .
\end{eqnarray}

\subsubsection{The FCS at the Toulouse point}
\label{FCSToulouse}

For realistic systems it is reasonable to assume $J_{z-}=0$. The
only remaining term which is still quartic in fermionic fields is
then zero at the so-called Toulouse point $J_{z+} = 2 \pi$
\cite{Toulouse,book}. In this situation the Hamiltonian is
quadratic in fermionic fields. The mixed GFs (\ref{KondoGFs}) are
related to the exact impurity GFs, $D_{bb}(t,t') = - i \langle T_C
b(t) b(t') \rangle_\lambda$ and to bare GFs (calculated for all
$J_i=0$) for the Majorana fields \cite{ourPRB} (notice that in the
present situation we have to double the applied voltage in
comparison to (\ref{bareGFs}), for details see \cite{ourPRB}),
\begin{eqnarray}                        \label{initialGFs}
g_{\xi \xi} = g_{\eta \eta} = \frac{i}{2} \left[ \begin{array}{cc}
 n_R + n_L - 1 &  n_R + n_L \\
 n_R + n_L - 2 &  n_R + n_L-1 \\
 \end{array}\right]
 \, \, \, \, \, , \, \, \, \, \, g_{\eta \xi} = \frac{n_L-n_R}{2} \left[
 \begin{array}{cc} 1 & 1 \\
 1 & 1 \\
 \end{array}\right] \, ,
\end{eqnarray}
in the following way,
\[
\begin{array}{ll}
G_{b\xi}(t,t')= &  i J_\perp \int\limits_{{\rm
C}}dt''D_{bb}(t,t'') \left\{ - \cos[\lambda (t'')/2]
g_{\xi\xi}(t''-t')+ \sin[\lambda(t'')/2]
 g_{\eta\xi}(t''-t')\right\} \;, \\
G_{b\eta}(t,t')= & i J_\perp \int\limits_{{\rm C}}dt''D_{bb}(t,t'')
\left\{-\cos[\lambda(t'')/2] g_{\xi\eta}(t''-t')
 + \sin[\lambda(t'')/2] g_{\eta\eta}(t''-t')\right\}\;.
\end{array}
\]
After the Keldysh disentanglement and using (\ref{initialGFs})
\begin{eqnarray}                   \label{Longformula}
\begin{array}{lll}
\frac{\partial}{\partial\lambda_-}{\cal U}(\lambda_-,\lambda_+)=
i\frac{J_\perp^2}{4} \int\frac{d\omega}{2\pi} \left\{
D_{bb}^{--}(\omega)(n_R - n_L) +  D_{bb}^{-+}(\omega)\left[ e^{i
\bar\lambda/2}(1-n_R) - e^{-i \bar\lambda/2} (1-n_L) \right]
\right\} \;.
\end{array}
\end{eqnarray}
Evaluation of the impurity GF is accomplished by the calculation
of the corresponding self-energy and inversion of the emerging
matrix $\hat{d}_0^{-1} - \hat{\Sigma}_K$ where in the absence of
the magnetic field $\Delta=0$ and $J_+=0$ (we discuss the general
case later)
\begin{eqnarray}   \label{SigmaforKondo}
 \hat{\Sigma}_K = \left[
\begin{array}{cc}
J_-^{2} g^{(0)--}_{\xi \xi} + J_\perp^2 g_{\xi\xi}^{--}\;\;\; & -
J_-^{2} g^{(0)-+}_{\xi \xi} - J_\perp^2 [c g_{\xi\xi}^{-+}- s
g_{\eta\xi}^{-+}] \\ - J_-^{2} g^{(0)+-}_{\xi \xi} - J_\perp^2 [c
g_{\xi\xi}^{+-} + s g_{\eta\xi}^{+-}]\;\;\; & J_-^{2}
g^{(0)++}_{\xi \xi} + J_\perp^2 g_{\xi\xi}^{++}
\end{array}
\right]\;,
\end{eqnarray}
where the super-script $(0)$ distinguishes the equilibrium GFs for
$V=0$ and $c=\cos[(\lambda_- - \lambda_+)/2]$, $s=\sin[(\lambda_-
- \lambda_+)/2]$ and $\hat{d}^{-1}$ is given in (\ref{cleardot})
with $\Delta_0=0$. Using (\ref{initialGFs}) and new definitions
$\Gamma_i = J_i^2/2$ we obtain
\begin{eqnarray}   \label{SigmaforKondo1} \nonumber
 \hat{\Sigma}_K = i \left[
\begin{array}{cc}
\Gamma_- (2 n_F - 1) + \Gamma_\perp (n_R + n_L -1) \;\;\; &
- \Gamma_- \, 2 n_F - \Gamma_\perp ( e^{i \bar\lambda/2} n_L + e^{- i \bar\lambda/2} n_R )
\\
\Gamma_- \, 2 (1-n_F) + \Gamma_\perp \left[ e^{i \bar\lambda/2}
(1-n_R) + e^{-i \bar\lambda/2} (1-n_L) \right] \;\;\; & \Gamma_-
(2 n_F - 1) + \Gamma_\perp ( n_R + n_L -1)
\end{array}
\right]\;.
\end{eqnarray}
Then the determinant
 \begin{eqnarray}   \nonumber
&-&\mbox{Det} \, (\hat{d}_{0}^{-1} - \hat{\Sigma}_K) =
\omega^2+(\Gamma_\perp+\Gamma_-)^2 +\Gamma_\perp^2
\left[n_L(1-n_R)(e^{i \bar\lambda}-1) +n_R(1-n_L)(e^{-i
\bar\lambda}-1)\right] \\ &+& 2\Gamma_- \Gamma_\perp
\left\{[n_F(1-n_R)+n_L(1-n_F)](e^{i \bar\lambda/2}-1)+
[n_F(1-n_L)+n_R(1-n_F)](e^{-i \bar\lambda/2}-1)\right\}\;.
\end{eqnarray}
The GFs of interest are then given by
\begin{eqnarray} \nonumber
 && \mbox{Det} \, (\hat{d}_{0}^{-1} - \hat{\Sigma}_K) \hat{D}_{bb} =
\\ \nonumber
 &&
\left[ \begin{array}{cc}
 - \omega +  i [\Gamma_- (2 n_F - 1) + \Gamma_\perp (n_R + n_L - 1)]
 &
  i \left[\Gamma_- \, 2 n_F + \Gamma_\perp (e^{i \bar\lambda/2} n_L +
e^{- i \bar\lambda/2} n_R)\right] \\
-  i \left\{\Gamma_- \, 2 (1-n_F) + \Gamma_\perp [e^{i
\bar\lambda/2} (1-n_R) + e^{- i \bar\lambda/2} (1-n_L)]\right\} &
\omega + i [\Gamma_- (2 n_F - 1) + \Gamma_\perp (n_R + n_L - 1)]
\end{array}
\right]
 \, .
\end{eqnarray}
 Inserting the calculated GFs into the fundamental relation (\ref{Longformula})
 results in
\begin{eqnarray}                          \label{fundrelationkondo}
 \frac{\partial}{\partial \lambda_-} {\cal U}(\lambda_\mp) = -i
 2 \Gamma_\perp \int \, \frac{d \omega}{2 \pi} \,
 \frac{I(\omega)}{\mbox{Det} \, (\hat{d}_{0}^{-1} - \hat{\Sigma}_K)}
\end{eqnarray}
with
\begin{eqnarray}  \nonumber
I(\omega) = \Gamma_\perp^2 \left[n_L(1-n_R)e^{i \bar\lambda}
-n_R(1-n_L)e^{-i \bar\lambda}\right] + 2\Gamma_- \Gamma_\perp
\left[n_F(1-n_R)e^{i \bar\lambda/2}- n_F(1-n_L)e^{-i
\bar\lambda/2}\right] \, .
\end{eqnarray}
To proceed, we split the $\omega$--integral in
Eq.~(\ref{fundrelationkondo}) into two parts for negative and
positive energies and change $\omega\to-\omega$ in the second
integral. In doing so observe that under this transformation
$n_F\to1-n_F$, $n_R\to 1-n_L$, and $n_L\to 1-n_R$. Therefore the
denominator stays invariant while the numerator changes as
\begin{eqnarray}               \nonumber
I(-\omega) &=& \Gamma_\perp^2 \left[n_L(1-n_R)e^{i\bar{\lambda}}
-n_R(1-n_L)e^{-i\bar{\lambda}}\right] + 2\Gamma_- \Gamma_\perp
\left[n_L(1-n_F)e^{i\bar{\lambda}/2}-
n_R(1-n_F)e^{-i\bar{\lambda}/2}\right]\;.
\end{eqnarray}
Eq.~(\ref{fundrelationkondo}) thus becomes
\begin{equation}\label{derivativebis}   \nonumber
\frac{\partial}{\partial \lambda_-} {\cal U}(\lambda_\mp) = -
\frac{1}{2} \int\limits_{0}^\infty \frac{d \omega}{2 \pi} \,
\frac{I(\omega)+I(-\omega)}{\mbox{Det} \, (\hat{d}_{0}^{-1} -
\hat{\Sigma}_K) } \;.
\end{equation}
Observe that, crucially, $\frac{\partial K}{\partial \lambda_-}=\frac{i}{2}
[I(\omega)+I(-\omega)]$
so that the $\lambda$--integration can be performed
as before.
The following exact formula for the statistics,
valid at finite temperatures, follows immediately:
\begin{eqnarray}             \label{finalkondo}
&~&\ln\chi(\lambda)={\cal T}  \int\limits_{0}^\infty \frac{d
\omega}{2 \pi} \ln \left\{1+T_2(\omega)
\left[n_L(1-n_R)(e^{2i\lambda}-1)
+n_R(1-n_L)(e^{-2i\lambda}-1)\right]\right.\label{stat2} \\
&~&\left.+T_1(\omega)\left[[n_F(1-n_R)+n_L(1-n_F)](e^{i\lambda}-1)+
[n_F(1-n_L)+n_R(1-n_F)](e^{-i\lambda}-1)\right]\right\},\nonumber
\end{eqnarray}
where the effective ``transmission coefficients'' (two of them now) are:
\begin{equation}\label{twotrans}
T_2(\omega) = \frac{\Gamma_\perp^2}{\omega^2 + (\Gamma_- +
\Gamma_\perp)^2}\;,\;\;\; T_1(\omega) = \frac{2\Gamma_-
\Gamma_\perp}{\omega^2 + (\Gamma_- + \Gamma_\perp)^2}\;.
\end{equation}
In the more general case of finite magnetic field
$\Delta$ and $\Gamma_+$ the
result (\ref{finalkondo}) is exactly the same up to the modified
transmission coefficients (derived in Appendix A),
\begin{eqnarray}                              \label{TC}
 T_2 &=& \frac{\Gamma_\perp^2 (\omega^2 + \Gamma_+^2)}{\left[ \omega^2 -
 \Delta^2
 - \Gamma_+(\Gamma_\perp + \Gamma_-)\right]^2 +
 \omega^2 (\Gamma_+ + \Gamma_-
 + \Gamma_\perp)^2} \, \, , \nonumber \\
 T_1 &=& \frac{2 \Gamma_\perp \Gamma_- (\omega^2+\Gamma_+^2) + 2 \Delta^2
 \Gamma_\perp \Gamma_+}{\left[ \omega^2 - \Delta^2
 - \Gamma_+(\Gamma_\perp + \Gamma_-)\right]^2 +
 \omega^2 (\Gamma_+ + \Gamma_-
 + \Gamma_\perp)^2} \, \, .
\end{eqnarray}
In fact, since the refermionised Hamiltonian describes local
scattering of \emph{non-interacting} (Majorana) particles, the result
(\ref{finalkondo}) can as well be derived using the approach
originally conceived by Levitov and Lesovik for systems with known
scattering matrix \cite{levitovlesovik}. For the corresponding
calculation see Appendix B.

Using the properties $n_F(1-n_R) + n_L(1-n_F) = n_L(1-n_R)(1+\exp[-V/T])$ and
$n_F(1-n_L) + n_R(1-n_F) = n_R(1-n_L)(1+\exp[V/T])$ we can rewrite the result
in the form
\begin{eqnarray}             \label{finalkondo1}
&~&\ln\chi(\lambda)={\cal T}  \int\limits_{0}^\infty
\frac{d \omega}{2 \pi} \ln \left\{1+
n_L(1-n_R)\left[  T_2(\omega) (e^{2i\lambda}-1) +
  T_1(\omega)(e^{i\lambda}-1)(1+e^{-V/T}) \right] \right.
 \\ \nonumber
&~& \left. + n_R(1-n_L)\left[  T_2(\omega) (e^{-2i\lambda}-1) +
  T_1(\omega)(e^{-i\lambda}-1)(1+e^{V/T}) \right]
\right\},\nonumber
\end{eqnarray}
\end{widetext}
 We first take a look
onto the $T=0$ situation, when $\exp(-|V|/T) \rightarrow 0$. In
that case one can reduce the generating function to the
Levitov-Lesovik formula (\ref{LLformula}) for a spinful system
\footnote{W.~Belzig, private communication.}
\begin{eqnarray}                  \label{reduction}
 \chi(\lambda) = \left[ 1 + T_{e}(e^{i \lambda}-1) \right]^N \, ,
\end{eqnarray}
where $T_e = \sqrt{T_2(0)}=\Gamma_\perp/(\Gamma_\perp +
\Gamma_-)$. Hence in the low temperature limit we obtain the
conventional binomial statistics for the charge transfer through
the dot. Needless to say this is in accordance with the binomial
theorem stated in the previous Section. However, the reduction
(\ref{reduction}) is not possible for finite temperatures and
voltages. To the best of our knowledge Eq.~(\ref{finalkondo}) is
the first exact result showing non-trivial statistics at finite
energy scales. It can be interpreted in terms of two distinct
tunnelling processes: (i) tunnelling of single electrons and (ii)
tunnelling of electron pairs with opposite spins. As has already
been realised in \cite{SH}, at least in the regime $T,V \ll
\Delta$ tunnelling of single electrons is energetically very
costly as it requires a spin-flip. A simultaneous tunnelling of
two electrons, which is described by the terms with $2 \lambda$
and $T_2(\omega)$, leaves the dot spin effectively untouched,
making that kind of process the dominant transport channel. In
zero field the finite voltage is known to act as effective
magnetic field \cite{SH} so that this tunnelling mechanism is
always present regardless of the precise value of $\Delta$.

In the low energy sector $\omega \ll V,T$ the integral of
(\ref{finalkondo1}) can be performed explicitly in the spirit of
Ref.~\cite{lll}, resulting in
\begin{equation}\label{smallVT}   \nonumber
 \chi(\lambda) = \exp\left[ {\cal T} \frac{T}{2 h} (u^2 - v^2)
 \right] \, ,
\end{equation}
where $v = V/T$ and
\begin{widetext}
\begin{eqnarray}\label{defu}   \nonumber
 \cosh (u) &-& \cosh(v) =
 T_1 [ \cos \lambda - 1 + \cosh(v + i \lambda) - \cosh v ]
 + T_2 [ \cosh(v + i 2 \lambda) - \cosh v ]
 \, .
\end{eqnarray}
In the limiting case $V \gg T$ we recover the result
(\ref{reduction}) while for $V \ll T$ we obtain
\begin{eqnarray}   \nonumber
 \chi(\lambda) = \exp\left( - {\cal T} \frac{T}{2 h} \lambda_*^2
 \right) \, ,
\end{eqnarray}
where $\sin^2 (\lambda_*/2 ) = 4 T_e \sin^2 (\lambda/2) \left[ 1 -
T_e \sin^2 (\lambda/2) \right]$. The full transport coefficient
$T_0$ as calculated in \cite{SH} turns out to be a {\it composite}
one and it is recovered from $T_{1,2}$ through a very simple
relation: $T_0 = T_2 + T_1/2$. We have evaluated the first and the
second cumulant of the Kondo FCS Eq.~(\ref{finalkondo}) which are
the same as calculated by SH at all $V$ and $T$ \cite{SH}. We
shall not reproduce these two cumulants here and concentrate
instead on new results.

First we would like to analyse the equilibrium statistics at
$V=0$. From (\ref{finalkondo}) it is obvious that as $n_L=n_R$ all
odd order cumulants are identically zero. Then for the even order
cumulants we obtain
\begin{eqnarray}
 \langle \delta q^2\rangle &=& {\cal T} \int_0^\infty \frac{d \omega}{2 \pi}
n_F (1-n_F) 4 (T_1 + 2 T_2) \, , \nonumber \\ \nonumber
 \langle \delta q^4\rangle  &=& {\cal T} \int_0^\infty \frac{d \omega}{2 \pi}
\left[ 4 (T_1 + 8 T_2)
 n_F (1-n_F) - 48 ( T_1 + 2 T_2)^2 n_F^2 (1-n_F)^2 \right] \, ,
 \nonumber \\ \nonumber
 \langle \delta q^6\rangle &=& {\cal T} \int_0^\infty \frac{d \omega}{2 \pi}
\Big[ 4 (T_1 + 32
 T_2) n_F (1-n_F) + 1920 ( T_1 + 2 T_2)^3 n_F^3 (1-n_F)^3
 \\ \nonumber
 &-& 240 ( T_1 + 2
 T_2)( T_1 + 8 T_2) n_F^2 (1-n_F)^2 \Big] \, .
\end{eqnarray}
As for finite $f(0)$ one obtains $\int_0^\infty d \omega f(\omega)
n_F^n (1-n_F)^n \approx a_n T f(0)$ with $a_1 = 1/2$, $a_2=1/12$,
$a_3=1/60$, $a_4=1/280$ etc. \emph{all} equilibrium cumulants are
linear in temperature in the low energy sector. The lowest order
cumulant is then the conventional thermal Johnson-Nyquist noise
$S_{JN} \approx 4 G_0 T_0 T$, where $G_0$ is the conductance
quantum and $T_0$ is the transmission coefficient of the dot at
$\omega=0$.

In the opposite limit of finite voltage and $T=0$ we obtain for the third
cumulant
\begin{eqnarray}\label{ctree}    \nonumber
\langle \delta q^3\rangle &=& {\cal T} \int_{0}^V
\frac{d \omega}{2 \pi} \,[T_1+8T_2-3(T_1+2T_2)(T_1+4T_2)
+ 2(T_1+2T_2)^3]\; .
\end{eqnarray}
This simplifies further in zero field:
\[
\langle \delta q^3\rangle =
\frac{{\cal T}}{2\pi} \, \left\{ 2 \Gamma_\perp
\mbox{tan}^{-1}[V/(\Gamma_\perp + \Gamma_-)] - \frac{2 V
\Gamma_\perp^2}{[(\Gamma_\perp + \Gamma_-)^2 + V^2]^2}
[(\Gamma_\perp+\Gamma_-)^2 + 2
\Gamma_-(\Gamma_\perp + \Gamma_-) + 3 V^2
]\right\} \, ,
\]
\end{widetext}
possessing the following limiting forms:
\begin{eqnarray}                \label{T0asympt}
  \langle \delta q^3\rangle_{V \rightarrow 0} &\approx& {\cal T}
  \, G_0
  \frac{2 \Gamma_\perp \Gamma_-(\Gamma_- - \Gamma_\perp)}
{(\Gamma_\perp + \Gamma_-)^3} \, V \, , \\ \nonumber
  \langle \delta q^3\rangle_{V \rightarrow \infty} &\approx&
  {\cal T} \,
 \pi \, G_0 \, \Gamma_\perp   \, .
\end{eqnarray}
At low voltages the cumulant is negative for $\Gamma_- < \Gamma_\perp$.
Generally, under these conditions the $n$-th cumulant
appears to possess $n-2$ zeroes as a function of $V$,
according to numerics. The
saturation value in the limit $V \rightarrow \infty$ is
independent of the coupling in the spin--flavour channel because
the fluctuations in the biased
conducting charge--flavour channel
are much more pronounced than those in
the spin--flavour channel, which experiences only relatively weak
equilibrium fluctuations.

\begin{figure}
\vspace*{1.0cm}
\psfrag{t}{${\cal T}$}
\epsfig{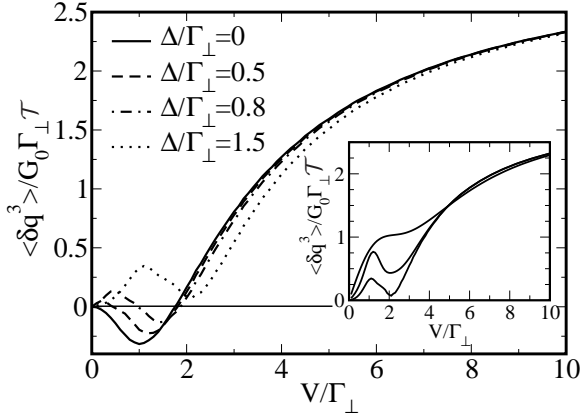}
\caption[]{\label{C3magnfield}
Zero temperature voltage dependence of the third cumulant
for different
magnetic field values and $\Gamma_\pm/\Gamma_\perp = 0.1$.
\emph{Inset:}
temperature evolution of the curve for $\Delta/\Gamma_\perp=1.5$ for
$T/\Gamma_\perp=0$, $0.2$ and $1.5$ (from bottom to top).}
\vspace*{0.0cm}
\end{figure}

For the general situation of arbitrary parameters, the cumulants
can be calculated numerically. The asymptotic value of the third
cumulant at high voltages, similarly to the findings of \cite{KT},
does not depend on temperature and is given by the result
(\ref{T0asympt}), see Fig.~1 of \cite{long}. In the opposite limit
of small $V$, $\langle \delta q^3 \rangle$ can be negative.
Sufficiently large coupling $\Gamma_-$ or magnetic field, see
Fig.~\ref{C3magnfield}, suppress this effect though.

According to the result of Ref.~\cite{reznikov},
as long as the distribution is binomial,
$\langle \delta q^3 \rangle/\langle \delta q\rangle=(e^*)^2$,
where $e^*$ is the effective charge of the current carriers.
This quantity is to be preferred to the
Schottky formula because of its weak temperature dependence.
Indeed we find numerically that the ratio
$\langle \delta q^3 \rangle/\langle \delta q\rangle$ in the present
problem is weakly temperature dependent
(it is flat and levels off to 1) in comparison to
$\langle \delta q^2 \rangle/\langle \delta q \rangle$.

\subsubsection{Corrections around the Toulouse point}
\label{corraround}

Thus far we dealt with a system which finds itself at one special
point in the parameter space, when $v_1=0$ and $J_{z-}=0$. While
the latter requirement is reasonable for realistic systems, the
former is quite artificial. It has been shown by means of RG
transformation procedure that at least in equilibrium the
operators, describing deviations from the Toulouse point are
irrelevant in the RG sense and do not influence the physics in the
low energy sector strongly. There is, however, no \emph{a priori} reason
why that should hold in a non-equilibrium situation. Therefore the
full analysis of the FCS must incorporate the investigation of the
statistics beyond the Toulouse restrictions. We first concentrate
on the situation of finite $v_1$. As was pointed out above, an
analytic solution in this situation is not possible. The only
option to progress is perturbation theory in $v_1$.

To access the generating function we still can use the fundamental
relation (\ref{Longformula}). As we have the complete knowledge of
all GFs with respect to $H'$, see (\ref{2chmajham}), the simplest
thing we can do is to calculate perturbative corrections to
$D_{bb}$ in the second order in $v_1$. They are given by
\begin{eqnarray}     \nonumber
 \hat{\bar{D}}_{bb} = \hat{D}_{bb} + v_1^2 \hat{D}_{bb} \, \hat{\Sigma}_b \, \hat{D}_{bb} =
 \hat{D}_{bb} + \delta  \hat{D}_{bb}  \, .
\end{eqnarray}
The correction to the adiabatic potential is then given by
\begin{widetext}
\begin{eqnarray}                       \label{ToulouseCorrection}
\nonumber
 \delta \left( \frac{\partial \cal U}{\partial \lambda_-} \right) = i
 \frac{\Gamma_\perp}{2} \int \, \frac{d \omega}{2 \pi} \, \left\{
 \delta D_{bb}^{--} (n_R - n_L) + \delta D_{bb}^{-+} \left[e^{i \bar\lambda/2} (1-n_R)
 - e^{-i \bar\lambda/2}(1-n_L ) \right] \right\} \, .
\end{eqnarray}
The self-energy matrix components are defined as
\begin{eqnarray}                \label{self-energy-definition}
\nonumber
 \Sigma^{ij}_b(\omega) = \int \frac{d \epsilon_1}{2 \pi} \, D^{ij}_{aa}
 (\omega-\epsilon_1) \,
 \int \frac{d \epsilon_2}{2 \pi} \,
 G^{ij}_s(\epsilon_2) G^{ji}_s(\epsilon_1 + \epsilon_2) \, ,
\end{eqnarray}
where $G_s(t,t') = - i \langle T_C \psi_s(t) \psi^\dag_s(t')
\rangle_\lambda$ is the GF of the spin sector fermion which is
free. Therefore at $T=0$ it is easily found to be
\begin{eqnarray}                        \label{freeGs} \nonumber
 G_s(\omega) = \left[ \begin{array}{cc}
                        -\frac{i}{2} \mbox{sgn}(\omega) & i \Theta(-\omega)
                        \\
                        - i \Theta(\omega)        &   -\frac{i}{2}
 \mbox{sgn}(\omega) \\
                      \end{array}
               \right] \, ,
\end{eqnarray}
In the $\Delta=0$ and $\Gamma_+=0$ case only $D_{aa}^{--(++)}=\pm
d_0 = \pm 1/\omega$ are non-zero, so that only $\Sigma_b^{--}$
needs to be calculated, resulting in
\[
 \Sigma_b^{--}(\omega) = - \Sigma_b^{++}(\omega) =
 (2 \pi)^{-2} \omega\left( \ln
 |\omega| - 1 \right) \, .
\]
It is an odd function of $\omega$ and vanishes in the infrared
limit as expected from RG arguments since it is generated by an
irrelevant operator. The corrections to the impurity GFs are then
\begin{eqnarray} \nonumber
 \delta D_{bb}^{--} =  v_1^2 \, \Sigma_b^{--} \left( D_{bb}^{--}
 D_{bb}^{--} - D_{bb}^{-+} D_{bb}^{+-} \right) \, \, \,  , \, \, \,
 \delta D_{bb}^{-+} = v_1^2 \, \Sigma_b^{--} D_{bb}^{-+} \left(
 D_{bb}^{--} - D_{bb}^{++} \right) \, .
\end{eqnarray}
Furthermore,
\begin{eqnarray}           \nonumber
  D_{bb}^{--}
 D_{bb}^{--} - D_{bb}^{-+} D_{bb}^{+-} &=& \mbox{Det}^{-1} \,
 (\hat{d}_0^{-1}-\hat{\Sigma}_K) + \frac{2 \omega \left\{ \omega +  i
 \left[ \Gamma_- (2 n_F -1) + \Gamma_\perp (n_R + n_L - 1)\right] \right\}}{\mbox{Det}^2 \,
 (\hat{d}_0^{-1}-\hat{\Sigma}_K)} \, ,
 \end{eqnarray}
and is an even function of $\omega$. After multiplication with the
self-energy, which is an odd function, and after an integration over all
frequencies we immediately see, that the off--Toulouse
correction to the first contribution in
 (\ref{Longformula}) is identically zero:
\begin{eqnarray}
\nonumber \
 D_{bb}^{-+} \left(  D_{bb}^{--} - D_{bb}^{++} \right) &=& -2 i
 \omega
 \frac{\Gamma_- \, 2 n_F + \Gamma_\perp ( e^{i \bar\lambda/2} n_L +
 e^{-i \bar\lambda/2} n_R )}{\mbox{Det}^2 \,
 (\hat{d}_0^{-1}-\hat{\Sigma}_K)} \, .
\end{eqnarray}
Therefore for the correction to the derivative of the adiabatic
potential we obtain
\begin{eqnarray}   \nonumber
 \delta \left( \frac{\partial {\cal U}}{\partial \lambda_-} \right) = - v_1^2
\Gamma_\perp \int \frac{d \omega}{2 \pi} \,
\frac{\Sigma_b^{--}}{\mbox{Det}^2 \,
 (\hat{d}_0^{-1}-\hat{\Sigma}_K)}
\omega \left\{ 2 \Gamma_- n_F[ e^{i \bar\lambda/2}(1-n_R) - e^{-i
\bar\lambda/2}(1-n_L) ]
 \right. \\ \nonumber \left.
 +  \Gamma_\perp
 [ e^{i \bar\lambda} n_L (1-n_R) - e^{-i \bar\lambda} n_R  (1-n_L)]
 \right\} \, ,
\end{eqnarray}
Comparing this result with (\ref{fundrelationkondo}) we conclude
that the effect of $v_1$ is the correction to the transmission
coefficients $\bar{T}_i = T_i + \delta T_i$ with
\begin{eqnarray}                    \label{prostayakorrektura}
 \nonumber
 \delta T_i(\omega)
 = \left( \frac{v_1}{2 \pi} \right)^2 \frac{\omega^2}{\omega^2+(\Gamma_- + \Gamma_\perp)^2} \left( \ln |\omega|
 -1 \right) T_i(\omega) \, ,
\end{eqnarray}
since schematically the structure of the correction is
\begin{equation}\label{correction} \nonumber
\frac{2T_2e^{2i\lambda}+T_1e^{i\lambda}}{1+
T_2(e^{2i\lambda}-1)+T_1(e^{i\lambda}-1)} +
\delta\frac{2T_2e^{2i\lambda}+T_1e^{i\lambda}}{[1+
T_2(e^{2i\lambda}-1)+T_1(e^{i\lambda}-1)]^2} \; ,
\end{equation}
which can be seen as the lowest order expansion of
\[
 \frac{2T_2(1+\delta) e^{2i\lambda}+T_1 (1+\delta) e^{i\lambda}}{1+
 T_2(1+\delta) (e^{2i\lambda}-1)+T_1 (1+\delta) (e^{i\lambda}-1)} \, .
\]
This correction vanishes for $\omega \rightarrow 0$, hence the trivialisation
(\ref{reduction}) still holds for the transmission coefficients away from
the Toulouse point.

It is not clear whether this picture is valid for $\Gamma_+$,
$\Delta \neq 0$. As in this situation the magnetic field couples
the $a$ and $b$ Majorana fields, the $D_{aa}$ correlation
functions have to be calculated from the Dyson equation with
respect to the self-energy $\Sigma_a^{ij} = (ij)\left[\Delta^2
D_{bb}^{ij}(\Delta=0) + \Gamma_+ g^{(0)ij}_{\eta \eta}\right]$. As
our ultimate goal is the behaviour of the generating function in
the limiting case of small $V$, we set $V=0$. We return to the
finite $V$ situation at the end of this Section. The resulting
$\Sigma_b$ can be best given in terms of the retarded component
$\Sigma_b^R$,
 \begin{eqnarray}                \label{sigmy}   \nonumber
 \Sigma_b^{--(++)} &=& \pm \, \mbox{Re} \Sigma_b^R - i \, (2n_F-1) \, \mbox{Im}
 \Sigma_b^R \, , \nonumber \\
 \Sigma_b^{-+} &=& i \, 2n_F  \, \mbox{Im} \Sigma_b^R \, , \\ \nonumber
 \Sigma_b^{+-} &=& - i \, 2 (1-n_F) \,  \mbox{Im} \Sigma_b^R \, ,
\end{eqnarray}
with
\begin{eqnarray}     \label{retardedcomponent}    \nonumber
 \Sigma_b^R &=& \frac{i}{(2 \pi)^2} \frac{1}{\sqrt{b^2-4 d}} \sum_{j=1,2} \,
 [\Omega^2_j + \Delta^2 - (\Gamma_-+\Gamma_\perp)^2]
  \left\{
 (\Omega_j-i \omega)
 \left[ \ln(\Omega_j-i \omega)-1\right]
 - \Omega_j (\ln \Omega_j - 1) \right\} \, ,
\end{eqnarray}
where $b=(\Gamma_-  + \Gamma_\perp)^2 + \Gamma_+^2 - 2 \Delta^2$, $d=[\Delta^2
  + \Gamma_+ (\Gamma_- + \Gamma_\perp)]^2$, and $\Omega^2_{1,2}=(b \pm
  \sqrt{b^2 - 4d})/2$. The expansion for small energies is different from that
  at $\Delta=0$,
\begin{eqnarray}         \nonumber
 \mbox{Re} \Sigma^R &\approx& - \frac{\omega}{(2 \pi)^2 \sqrt{b^2-4 d}}
 \sum_{j=1,2} \,
 [\Omega^2_j + \Delta^2 - (\Gamma_-+\Gamma_\perp)^2] \ln \Omega_j
 \nonumber \\ \nonumber
 \mbox{Im} \Sigma^R &\approx& - i \frac{\omega^2}{(2 \pi)^2 2 \sqrt{b^2-4 d}}
 \sum_{j=1,2} \, [\Omega^2_j + \Delta^2 - (\Gamma_-+\Gamma_\perp)^2] \frac{\ln
 \Omega_j}{\Omega_j} \, .
\end{eqnarray}
The correction to the time ordered part is,
\begin{eqnarray}
 \delta D_{bb}^{--} &=&
v_1^2 \, \left(  D_{bb}^{--} \Sigma_b^{--}
 D_{bb}^{--} + D_{bb}^{-+} \Sigma_b^{++} D_{bb}^{+-} + D_{bb}^{-+} \Sigma_b^{+-}
 D_{bb}^{--} + D_{bb}^{--} \Sigma_b^{-+} D_{bb}^{+-} \right) \,
 \nonumber \\ \nonumber
 &=& v_1^2 \left( D_{bb}^{--}D_{bb}^{--}
- D_{bb}^{-+} D_{bb}^{+-} \right) \mbox{Re}
 \Sigma_b^R + i v_1^2 \, \mbox{Im}\Sigma_b^R
 \, {\cal F}_1(\omega)\;,
\end{eqnarray}
with
\begin{eqnarray}              \nonumber
 {\cal F}_1(\omega)
 = -(2n_F-1) \left( D_{bb}^{--} D_{bb}^{--} + D_{bb}^{-+}D_{bb}^{+-}
 \right)
 - 2 (1-n_F) D_{bb}^{--} D_{bb}^{-+} + 2 n_F
 D_{bb}^{--} D_{bb}^{+-}  \, .
\end{eqnarray}
This function is an odd function of $\omega$ while
$D_{bb}^{--}D_{bb}^{--} - D_{bb}^{-+} D_{bb}^{+-}$ is even. Taking
into account the symmetry properties of the self-energy we
conclude that the whole contribution to the generating function
stemming from $\delta D_{bb}^{--}$ vanishes. The other component
can be written down in the similar way,
\begin{eqnarray}
 \delta D_{bb}^{-+}
&=& \nonumber v_1^2 \, \left(  D_{bb}^{--} \Sigma_b^{--}
 D_{bb}^{-+} + D_{bb}^{-+} \Sigma_b^{++} D_{bb}^{++} + D_{bb}^{-+} \Sigma_b^{+-}
 D_{bb}^{-+} + D_{bb}^{--} \Sigma_b^{-+} D_{bb}^{++} \right) \\ \nonumber
 &=& v_1^2 D_{bb}^{-+}(D_{bb}^{--}-D_{bb}^{++}) \mbox{Re} \Sigma_b^R
 + \, i v_1^2 \,
 \mbox{Im} \Sigma_b^R  {\cal F}_2(\omega)\, ,
\end{eqnarray}
where we have introduced
\begin{eqnarray}            \label{F2}
{\cal F}_2(\omega) &=& (1-2n_F) D_{bb}^{-+} (D_{bb}^{--} +
D_{bb}^{++}) + 2 (1-n_F) D_{bb}^{-+}D_{bb}^{-+} - 2 n_F
D_{bb}^{--} D_{bb}^{++} \, .
\end{eqnarray}
The analysis of the contribution arising from Re$\Sigma_b^R$ can
be done in the same way as before as it has exactly the same
structure. The substitution (\ref{substitution}) still can be
applied and one immediately recognises that it only leads to the
renormalisation of the transmission coefficients,
\begin{eqnarray}                     \label{corrtoT12}
 \delta T_i(\omega) = v_1^2 \mbox{Re} \Sigma_b^R
 \frac{ \omega (\omega^2 + \Gamma_+^2 - \Delta^2)}{
 \left[ \omega^2 - \Delta^2 -
 \Gamma_+(\Gamma_\perp + \Gamma_-)\right]^2 + \omega^2 (\Gamma_- + \Gamma_+
 +\Gamma_\perp)^2} T_i(\omega) \, .
\end{eqnarray}
Re$\Sigma_b^R$ is itself linear in $\omega$ in the low energy
sector, that is why the corrections to the transmission
coefficients (\ref{corrtoT12}) vanish at low energies. Now we turn
to the contribution of Im$\Sigma_b^R$. The first term in
(\ref{F2}) can be shown to produce renormalisation of the
transmission coefficient similar to (\ref{corrtoT12}),
\begin{eqnarray}                     \label{corrtoT12prime}
\nonumber
 \delta T_i(\omega) = v_1^2 \mbox{Im} \Sigma_b^R
 \frac{\Gamma_- \left[ \Gamma_- (\omega^2 + \Gamma_+^2) + \Delta^2 \Gamma_+ \right]}{\left[ \omega^2 - \Delta^2 -
 \Gamma_+(\Gamma_\perp + \Gamma_-)\right]^2 + \omega^2 (\Gamma_- + \Gamma_+
 +\Gamma_\perp)^2} T_i(\omega) \, .
\end{eqnarray}
Although the two remaining terms of (\ref{F2}) cannot be reduced
to renormalisation of the transmission coefficients in a simple
way, their contribution to the derivative of the adiabatic
potential can be evaluated directly,
\begin{eqnarray}            \nonumber
\delta_I \left( \frac{\partial {\cal U}}{ \partial \lambda_-}
\right) = v^2 \frac{\Gamma_\perp}{2} \int_0^V \, \frac{d \omega}{2
\pi} \, \mbox{Im} \Sigma_b^R e^{i \bar\lambda/2} \mbox{Det}^{-2}
\,
 (\hat{d}_0^{-1}-\hat{\Sigma}_K) \left[ \left(\omega - \frac{\Delta^2 \omega}{\omega^2 +
 \Gamma_+^2}\right)^2 + \left( \Gamma_- +
 \frac{\Delta^2 \Gamma_+}{\omega^2 + \Gamma_+^2} \right)^2 +
 \Gamma_\perp^2 e^{i \bar\lambda} \right] \, .
 \end{eqnarray}
Taking into account that the leading behaviour of the imaginary
part of the self-energy is $\sim \omega^{2}$ for small energies
one immediately verifies, that the above correction is cubic in
the applied voltage and therefore leads to qualitatively the same
picture as the renormalisation of the transmission coefficients.
For that particular evaluation we
used the equilibrium self-energy $\hat{\Sigma}_b$. Nevertheless,
after a lengthy but straightforward
calculation, we find that the same conclusion is still valid for the proper
non-equilibrium one as the corresponding corrections to
$\hat{\Sigma}_b$ is of exactly the same order in $V$ and $\omega$.
Thus at least in the low energy sector the predictions of
Sec.~\ref{FCSToulouse} remain valid beyond the Toulouse point.

\subsubsection{Resonant level problem in Luttinger liquids}
\label{RLLL}

We now briefly turn to the $g=1/2$ RL set--up. This set--up has
caused much interest recently, see Ref.~\cite{ourPRL} and
references therein. The Hamiltonian now is
\begin{equation}\label{12ham}      \nonumber
H=H_0+(\gamma_L \, \psi_L d^\dagger + \gamma_R \,
d\psi_R^\dagger+{\rm H.c})+ \Delta d^\dagger d+H_C\;,
\end{equation}
where $H_0$ stands for two biased Luttinger liquids (LLs), $d$ is
the electron operator on the dot, $\gamma_{R(L)}$ are the
tunnelling amplitudes to $R(L)$ electrode and $H_C$ is an
electrostatic interaction (see also \cite{ourPRL}),
\begin{eqnarray} \nonumber
 H_C =  \lambda_C d^\dag d \, \sum_i \, \psi_i^\dag(0) \psi_i(0)
 \,.
\end{eqnarray}
The contacting electrodes are supposed to be one-dimensional
half-infinite electron systems. We model them by chiral fermions
living in an infinite system: the negative half-axis then
describes the particles moving towards the boundary, while the
positive half-axis carries electrons moving away from the end of
the system. In the bosonic representation $H_0[\psi_i]$ are
diagonal even in presence of interactions (for a recent review see
e.~g. \cite{book}; we set the renormalised Fermi velocity
$v=v_F/g=1$, the bare velocity being $v_F$):
\begin{eqnarray}                 \label{Hi}       \nonumber
 H_0[\psi_i] = (4 \pi)^{-1} \int \, dx \, [\partial_x \phi_i(x)]^2.
\end{eqnarray}
Here the phase fields $\phi_i(x)$ describe the slow varying
spatial component of the electron density (plasmons),
\begin{eqnarray} \nonumber
\psi^\dag_i(x) \psi_i(x) = \partial_x \phi_i(x)/2 \pi \sqrt{g} \,
.
\end{eqnarray}
The electron field operator at the boundary is given by
\footnote{Strictly speaking $\psi(x=0)=0$, so we assume that the
tunnelling takes place at the second last site of the
corresponding lattice model, at $x=\pm a_0$. },
\begin{eqnarray}                           \label{psioperator}
\nonumber
 \psi_i(0) = e^{i \phi_i(0)/\sqrt{g}}/\sqrt{2 \pi a_0} \, ,
\end{eqnarray}
where $a_0$ is the lattice constant of the underlying lattice
model. Here $g$ is the conventional LL parameter (coupling
constant) connected to the bare interaction strength $U$ via
$g=(1+U/\pi v_F)^{-1/2}$ \cite{book,kanefisher}. In the chiral
formulation the bias voltage amounts to a difference in the
densities of the incoming particles in both channels far away from
the constriction \cite{eggergrabert,grabert}. The current is then
proportional to the difference between the densities of incoming
and outgoing particles within each channel.

Construction of the operator (\ref{Tlambdaoperator}) is
unproblematic and leads to
\begin{eqnarray}
T_\lambda = \gamma_L (d^\dag \psi_L e^{i \lambda/4} + \psi_L^\dag
d e^{-i \lambda/4} ) \nonumber &+&  \gamma_R (d^\dag \psi_R e^{-i
\lambda/4} + \psi_R^\dag d e^{i\lambda/4} ) \, ,
\end{eqnarray}
where, contrary to the Anderson impurity calculation, we choose to
build in the counting field in a symmetric manner for the reasons
which will become clear later. After the Emery--Kivelson rotation,
refermionization to new fermions $\psi$ and after the introduction
of the Majorana components as in (\ref{Majoranashow})
\cite{EK,ourPRB} we find
\begin{eqnarray}                   \label{Tlambdag0.5} \nonumber
T_\lambda &=& \left[ e^{i \lambda/4} ( \gamma_L d^\dag \psi +
\gamma_R \psi d) + e^{-i
    \lambda/4} (\gamma_L d^\dag \psi^\dag + \gamma_R \psi^\dag d)\right]
\nonumber \\
 &=&
-i \gamma_+ \, b \left[ \cos(\lambda/4) \xi - \sin(\lambda/4) \eta
\right] + i \gamma_- \, a \left[ \sin(\lambda/4) \xi -
\cos(\lambda/4) \eta \right] \;.
\end{eqnarray}
where $\gamma_\pm = \gamma_L \pm \gamma_R$. In case of the
symmetric coupling $\gamma_-=0$ the corresponding $T_\lambda$ has
exactly the same shape as (\ref{TlambdaKondo}). In fact we find
the same set of equations as for the Kondo dot,
Eq.~(\ref{2chmajham}) and Eq.~(\ref{TlambdaKondo}), but with
$\lambda\to\lambda/2$ and $J_\perp=\gamma_+$, $J_\pm=0$.
Consequently, the FCS is given by a modification of the
Levitov--Lesovik formula (\ref{LLformula}):
\begin{equation}\label{12stat}
\chi_{1/2}(\lambda)= \chi_0 (\lambda;2V;\{T_\Delta(\omega)\}) \;,
\end{equation}
with the effective transmission coefficient
$T_{\Delta}(\omega)=4\gamma^4\omega^2/
[4\gamma^4\omega^2+(\omega^2-\Delta^2)^2]$ of the RL set-up in the
symmetric case \cite{ourPRL}. All the cumulants are thus
obtainable from those of the non-interacting statistics
Eq.~(\ref{LLformula}).

The $\Delta=0$ RL set--up is equivalent to the model of direct
tunnelling between two $g=2$ LLs \cite{ourPRB}. The latter model
is connected by the strong to weak coupling ($1/g\to g$) {\it
duality} argument to the $g=1/2$ Kane and Fisher model
\cite{kanefisher,saleurduality}, which is, in turn, equivalent to
the CB set--up studied in \cite{AM,KT} (for a more general case of
arbitrary interaction strength see also \cite{weisssaleur,safi}).
Therefore their FCS must be related to our Eq.~(\ref{12stat}) at
$\Delta=0$ by means of the transformation: $T_0\to 1-T_0$ and
$V\to V/2$. Indeed after some algebraic manipulation with Eq.~(12)
of \cite{KT}, for details see Appendix C, we find that the FCS for
the CB set--up can be re--written as:
\begin{equation}\label{CBstat}     \nonumber
\chi_{CB}(\lambda)= \chi_0 (-\lambda;V;\{1-T_0(\omega)\}) \;.
\end{equation}

For the asymmetric coupling $\gamma_- \neq 0$ the problem cannot
be mapped onto the Kondo dot any more. The corresponding
calculation is nevertheless straightforward and is presented in
Appendix D. There is no fundamental difference in the result up to
the more involved transmission coefficient, which has already been
derived for the case of the non-linear $I-V$ in \cite{ourPRL}.

\section{Conclusions}
\label{conclusions}

To conclude, we present a detailed study of the charge transfer
statistics through the Anderson impurity model. We find an
expression for the exact generating function in terms of the
impurity self--energy calculated in the presence of the measuring
field $\lambda$: Eq.~(\ref{fi}). Based on this formula we conclude
that $T=0$ linear response statistics is {\it universal} and
binomial for the AIM and similar models: we call this fact {\it
binomial theorem}. The only effect of correlations is to define an
effective transmission coefficient. For the symmetric AIM, for
example, there is a perfect transmission and no fluctuations of
the current at all in this case.

In the search for non-trivial interaction effects one has, therefore,
to go to higher values of $T$ and $V$. To this end we have calculated the
exact FCS distribution function in the Toulouse limit (Kondo
regime): it is given by Eq.~(\ref{finalkondo}). This formula
uncovers rather profound, if model dependent, consequences of
correlations: there are two distinct tunnelling processes ($T_1$
and $T_2$), that of single electrons and electron pairs with
opposite spin. The latter process is, in fact, dominant in zero
field. The structure of higher moments is also determined by these
two processes as discussed in detail in the main text. At $T=0$
linear response all this rich physics is masked by
Eq.~(\ref{finalkondo}) collapsing to the universal binomial
distribution. We checked this universality by extensively studying
corrections to the distribution function due to departures from
the Toulouse limit.

We close by outlining some possible directions for future developments.
Formula (\ref{fi}) could be used to develop Fermi liquid theory for
the noise and possibly higher moments. Perhaps more importantly, the
ideas of this paper could be applied to models with many conduction
channels, where one would expect some equivalent of the binomial
theorem to hold, as seems to be compatible with recent experiments
\cite{bomze}.

\acknowledgements

We wish to thank H.~Saleur, H.~Grabert, K.~Sch\"{o}nhammer,
A.~Tsvelik, A.~Chitov, Y.~Adamov, F.~Siano
and R.~Egger for many inspiring discussions. We would also like to
thank W.~Belzig for illuminating discussions and for pointing out
to us the reduction: Eq.~(\ref{reduction}). Part of this work was
done during AOG's visits to the Brookhaven National Laboratory and
to the University of G\"{o}ttingen, the hospitality is kindly
acknowledged. The AK's financial support has been provided  by the
EU RTN programme, by the DFG and by the Landesstiftung
Baden-W\"urttemberg. AK is Feodor Lynen fellow of the Alexander
von Humboldt foundation.

\section*{Appendix A}

The calculation of the impurity GF in finite magnetic field
$\Delta$ and $J_+ \neq 0$ is accomplished again by inversion of
$\hat{d}_0^{-1} - \hat{\Sigma}_\Delta$ with
\begin{eqnarray}                  \nonumber
 \hat{\Sigma}_\Delta= \left[
 \begin{array}{cc}
 \Delta^2 D_{aa}^{--} + \Sigma^{--}_K\;\;\; & - \Delta^2 D_{aa}^{-+} + \Sigma^{-+}_K \\
 -\Delta^2 D_{aa}^{+-} + \Sigma^{+-}_K \;\;\; & \Delta^2 D_{aa}^{++} +
 \Sigma^{++}_K
 \end{array}
 \right]\; ,
\end{eqnarray}
where $\hat{\Sigma}_K$ is given in (\ref{SigmaforKondo}).
$\hat{D}_{aa}$ has to be evaluated with respect to the Hamiltonian
\begin{eqnarray}               \label{hamgamma+} \nonumber
 H_+ = H_0[\eta_f] - i J_+ \, a \, \eta_f \, .
\end{eqnarray}
A relatively simple calculation yields
\begin{eqnarray}                   \label{daawithgamma+}
\nonumber
  D_{aa}(\omega) = \frac{1}{\omega^2 + \Gamma_+^2} \left[
\begin{array}{cc}
 \omega + i \Gamma_+ (2 n_F - 1) & i 2\Gamma_+ n_F \\
 - i 2 \Gamma_- (1-n_F) &  - \omega + i \Gamma_+ (2 n_F - 1) \\
\end{array} \right] \, .
\end{eqnarray}
From now on the calculation can be performed in exactly the same
way as before. However, writing down explicitly the expression for
$\hat{\Sigma}_\Delta$ one observes, that it can be constructed
from the corresponding $\hat{\Sigma}_K$ via trivial substitution
\begin{eqnarray}                      \label{substitution}
 \omega \rightarrow \omega - \frac{\Delta^2 \omega}{\omega^2 +
 \Gamma_+^2} \, \, \, , \, \, \, \Gamma_- \rightarrow \Gamma_- +
 \frac{\Delta^2 \Gamma_+}{\omega^2 + \Gamma_+^2} \, .
\end{eqnarray}
This can be used to obtain the transmission coefficients
(\ref{TC}) from the ones given by (\ref{twotrans}).

\section*{Appendix B}

For systems with known scattering matrix $s_{\alpha \beta, m n}$
between terminals $\alpha, \beta$ and channels $m$ and $n$, there
is a ready formula for the FCS generating function, derived in
\cite{levitovlesovik},
\begin{eqnarray}                      \label{LLmainresult}
\ln \chi(\lambda) = \frac{\cal T}{2 \pi} \int \, d \omega \, \ln
 \mbox{Det} \, \left[ 1 + \hat{f}(\omega) \, \left( \hat{s}^\dag
 \, \widetilde{s} \, - 1 \right) \right] \, ,
\end{eqnarray}
where $ \widetilde{s}_{\alpha \beta, m n} = e^{i (\lambda_\alpha -
\lambda_\beta) } s_{\alpha \beta, m n}$, $\lambda_{\alpha, \beta}$
being the fields counting the particles in the respective
terminals. $\hat{f}(\omega) = \delta_{m n} \, \delta_{\alpha
\beta} \, f_\alpha(\omega)$ is diagonal in both channel ($m,n$)
and terminal ($\alpha,\beta$) indices and describes the energy
distribution function in the respective terminal. In the simplest
situation, when $\Delta=0$ and $ J_+=0$, we have four terminals
with one channel in each of them.  The scattering part of the
Hamiltonian is $H_I = J_\perp (\psi^\dag - \psi) \, b + J_-
(\psi_s^\dag - \psi_s) \, b$, see (\ref{KondoHam})(for simplicity
we ignore the unimportant numerical prefactors). The equations of
motion (EoMs) for the participating operators read
\begin{eqnarray}
 i \partial_t \psi_s &=& - i \partial_x \psi_s + J_- b \delta(x) \,
 , \nonumber \\
 i \partial_t \psi &=& - i \partial_x \psi + J_\perp b \delta(x)
 \, , \nonumber \\ \nonumber
 i \partial_t b &=& J_\perp (\psi^\dag - \psi) + J_- (\psi_s^\dag
 - \psi_s) \, .
\end{eqnarray}
Integrating the first equation over time and then around the point
$x=0$ we obtain
\begin{eqnarray}   \nonumber
 i \left[ \psi_s(0^+) - \psi_s(0^-) \right] = J_- b \, .
\end{eqnarray}
Acting with $i \partial_t$ from the left and using the EoM for the
$b$ Majorana one obtains,
\begin{eqnarray}
 - \partial_t \left[ \psi_s(0^+) - \psi_s(0^-) \right] &=& J_-
 J_\perp (\psi^\dag - \psi) + J_-^2 (\psi_s^\dag - \psi_s) \, ,
 \nonumber \\ \nonumber
- \partial_t \left[ \psi(0^+) - \psi(0^-) \right] &=& J_-
 J_\perp (\psi_s^\dag - \psi_s) + J_\perp^2 (\psi^\dag - \psi) \,
 ,
\end{eqnarray}
where the last equation is obtained by symmetry. Now we employ the
plain wave decomposition similar to that used in
\cite{wen,eggergrabert,ourPRL},
\begin{eqnarray} \nonumber
 \psi(x) = \int \, \frac{d k}{2 \pi} \, e^{i k (x-t)} \, \left\{
 \begin{array}{cc}
  a_k & \mbox{for} \, x<0 \\
  b_k & \mbox{for} \, x>0
 \end{array}
 \right.
 \, \, \, , \, \, \,
\psi_s(x) = \int \, \frac{d k}{2 \pi} \, e^{i k (x-t)} \, \left\{
 \begin{array}{cc}
  c_k & \mbox{for} \, x<0 \\
  d_k & \mbox{for} \, x>0
 \end{array}
 \right.
 \, .
\end{eqnarray}
Since the dispersion relation of both fermion species is trivial,
$\omega = k$, we can use $\omega$ both for momentum and energy.
Employing the regularisation scheme $\psi_i = [\psi(0^+) +
\psi(0^-)]/2$ we obtain
\begin{eqnarray}
 - i \omega \left(d_\omega - c_\omega \right) &=& \frac{J_\perp J_-}{2} \left(
 a_{-\omega}^\dag + b_{-\omega}^\dag - a_\omega - b_\omega \right)
 + \frac{J_-^2}{2} \left( c_{-\omega}^\dag + d_{-\omega}^\dag -
 c_\omega - d_\omega \right) \, , \nonumber \\ \nonumber
- i \omega \left( b_\omega - a_\omega \right) &=& \frac{J_\perp
J_-}{2} \left(
 c_{-\omega}^\dag + d_{-\omega}^\dag - c_\omega - d_\omega \right)
 + \frac{J_\perp^2}{2} \left( a_{-\omega}^\dag + b_{-\omega}^\dag -
 a_\omega - b_\omega \right) \, .
\end{eqnarray}
Comparing these relations with their adjunct at $-\omega$ we
identify that
\begin{eqnarray}
 b_{-\omega}^\dag &=& a_{-\omega}^\dag - b_\omega + a_\omega \, ,
 \nonumber \\ \nonumber
 d_{-\omega}^\dag &=& c_{-\omega}^\dag - d_\omega + c_\omega \, ,
\end{eqnarray}
and that $b_\omega - a_\omega = (d_\omega - c_\omega)\, J_\perp
/J_-$. Using these expressions we can find both $b_\omega$ and
$d_\omega$ as functions of $a_\omega$ and $c_\omega$, e.~g.
\begin{eqnarray} \nonumber
 b_{\omega} = \frac{1}{\omega + i (\Gamma_\perp + \Gamma_-)}
 \left[ (\omega + i \Gamma_-) \, a_\omega + i \Gamma_\perp \,
 a_{-\omega}^\dag - i
 J_\perp J_- \, c_\omega + i J_\perp J_- \,
 c_{-\omega}^\dag   \right]\, .
\end{eqnarray}
That leads to the following scattering matrix,
\begin{eqnarray}    \nonumber
 \left( \begin{array}{c}
 b_\omega \
 b_{-\omega}^\dag \\
 d_\omega \\
 d_{-\omega}^\dag \end{array} \right) = s \,  \left( \begin{array}{c}
 a_\omega \\
 a_{-\omega}^\dag \\
 c_\omega \\
 c_{-\omega}^\dag \end{array} \right) = \frac{1}{\omega + i
 (\Gamma_\perp + \Gamma_-)} \left( \begin{array}{cccc}
 \omega + i \Gamma_- & i \Gamma_\perp & - i J_\perp J_- & i
 J_\perp J_- \\
 i \Gamma_\perp & \omega + i \Gamma_- & i J_\perp J_- & - i
 J_\perp J_- \\
 - i J_\perp J_- & i J_\perp J_- & \omega + i \Gamma_\perp & i
 \Gamma_-  \\
 i J_\perp J_- & - i J_\perp J_- & i \Gamma_- & \omega + i
 \Gamma_\perp
 \end{array} \right) \,
 \left( \begin{array}{c}
 a_\omega \\
 a_{-\omega}^\dag \\
 c_\omega \\
 c_{-\omega}^\dag \end{array} \right) \, .
\end{eqnarray}
The actual charge transport through the system is conveyed by the
charge flavour channel, e.~g. by scattering of $\psi$ fermions
across the constriction. The physical picture is similar to that
discussed in \cite{ourPRL}: the incoming particles -- chiral
fermions in terminal $1$, which are described by $a_k$ operators
and which have chemical potential $\mu_1=V$ -- are transferred
into all other terminals $2$-$4$ ($b_k$, $c_k$ and $d_k$
operators), which are unbiased $\mu_{2,3,4}=0$, that is why we
have to set $\hat{f} = \mbox{diag}(n_L, n_F, n_F, n_F)$. Then
$\lambda_1 = \lambda$ counts particles which leave channel $1$.
However, the very same fermion reappears in the channel $2$. Since
$1$ and $2$ are physically one lead we have $\lambda_2 = -
\lambda$. We are not interested in change of particle numbers in
the other channels, that is why $\lambda_{3,4}=0$ \footnote{We
would like to address the question of counting statistics for
transferred spin in a future publication.}. Therefore the matrix
$\widetilde{s}$ is given by
\begin{eqnarray} \nonumber
 \widetilde{s} =  \frac{1}{\omega + i (\Gamma_\perp + \Gamma_-)} \left( \begin{array}{cccc}
 \omega + i \Gamma_- & i \Gamma_\perp \, e^{-i 2 \lambda} & - i J_\perp J_- \, e^{-i \lambda} & i
 J_\perp J_- \, e^{-i \lambda} \\
 i \Gamma_\perp \, e^{i 2 \lambda} & \omega + i \Gamma_- & i J_\perp J_- e^{i \lambda} & - i
 J_\perp J_- e^{i \lambda} \\
 - i J_\perp J_- \, e^{i \lambda}& i J_\perp J_- e^{-i \lambda}& \omega + i \Gamma_\perp & i
 \Gamma_- \\
  i J_\perp J_- \, e^{i \lambda}& - i J_\perp J_- e^{-i \lambda}& i \Gamma_- & \omega + i
 \Gamma_\perp
 \end{array} \right) \, .
\end{eqnarray}
Plugging these relations into (\ref{LLmainresult}), folding the
integration over energy to the domain $[0,\infty)$ and using the
properties $n_F(-\omega) = 1-n_F(\omega)$ and $n_L(-\omega) = 1 -
n_R(\omega)$ immediately leads then to the result
(\ref{finalkondo}).

\section*{Appendix C}

Here we establish the relation between our findings and the result
of Kindermann and Trauzettel (KT) calculation \cite{KT}. Let us
consider Eq.~(12) of Ref.~\cite{KT},
\begin{eqnarray}\label{KT}        \nonumber
\ln\chi_{KT}(\lambda)=\frac{\cal T}{4\pi}(-iV\lambda-T^2\lambda^2)
+  {\cal T}
\int\limits_0^\infty\frac{d\omega}{2\pi}\ln\left\{1+T_0(\omega)
[f^+(1-f^-)(e^{i\lambda}-1)+f^-(1-f^+)(e^{-i\lambda}-1)]\right\}\;,\nonumber
\end{eqnarray}
The $f$-functions are given by
\[
f^\pm=\frac{n_{L/R}}{(1-n_{L/R})e^{\pm i\lambda}+n_{L/R}}\;,
\]
where $n_{R,L}(\omega)=n_F(\omega\pm V/2)$. Obviously
\[
1-f^\pm=\frac{(1-n_{L/R})e^{\pm i\lambda}}{(1-n_{L/R})e^{\pm
i\lambda}+n_{L/R}}\;,
\]
Taking into account that
$T_0(\omega)=4\gamma^4/(\omega^2+4\gamma^2)$, the identification
with the KT's impurity strength is $2\gamma=T_B$. In order to
proceed we define the object
\begin{eqnarray}\label{KTalpha}
\ln\chi_{KT}(\lambda;\alpha) = \frac{\cal
  T}{4\pi}(-iV\lambda-T^2\lambda^2) +
{\cal T}
\int\limits_0^\infty\frac{d\omega}{2\pi}\ln\left\{1+\alpha
T_0(\omega)
[f^+(1-f^-)(e^{i\lambda}-1)+f^-(1-f^+)(e^{-i\lambda}-1)]\right\}\;,
\end{eqnarray}
where $\alpha$ is a parameter. The derivative of (\ref{KTalpha})
with respect to this parameter
\[
\frac{\partial\ln\chi_{KT}(\lambda;\alpha)}{\partial\alpha}= {\cal
T} \int\limits_0^\infty\frac{d\omega}{2\pi} \frac{T_0(\omega)
[f^+(1-f^-)(e^{i\lambda}-1)+f^-(1-f^+)(e^{-i\lambda}-1)]}{1+
\alpha
T_0(\omega)[f^+(1-f^-)(e^{i\lambda}-1)+f^-(1-f^+)(e^{-i\lambda}-1)]}\;.
\]
Substituting explicit
expressions for the $f$--functions into the long fraction gives
\[
-\frac{T_0(\omega)[n_L(1-n_R)(e^{-i\lambda}-1)+n_R(1-n_L)(e^{i\lambda}-1)
]}{[(1-n_L)e^{i\lambda}+n_L][(1-n_L)e^{-i\lambda}+n_R]- \alpha
T_0(\omega)[n_L(1-n_R)(e^{-i\lambda}-1)+n_R(1-n_L)(e^{i\lambda}-1)]}\;.
\]
After simple algebra this simplifies as
\[
-\frac{T_0(\omega)[n_L(1-n_R)(e^{-i\lambda}-1)+n_R(1-n_L)(e^{i\lambda}-1)
]}{1+[1- \alpha
T_0(\omega)][n_L(1-n_R)(e^{-i\lambda}-1)+n_R(1-n_L)(e^{i\lambda}-1)]}\;.
\]
Integrating with respect to $\alpha$ therefore results in
\[
\ln\chi_{KT}(\lambda;\alpha)= {\cal T}
\int\limits_0^\infty\frac{d\omega}{2\pi}\ln\left\{1+[1- \alpha
T_0(\omega)][n_L(1-n_R)(e^{-i\lambda}-1)+n_R(1-n_L)(e^{i\lambda}-1)]
\right\}+C\;.
\]
To fix the constant $C$, evaluate the above integral at
$\alpha=0$, see also \cite{lll}
\[
{\cal T} \int\limits_0^\infty\frac{d\omega}{2\pi}\ln[1+
n_L(1-n_R)(e^{-i\lambda}-1)+n_R(1-n_L)(e^{i\lambda}-1)]=
\frac{\cal T}{4\pi}(-iV\lambda-T^2\lambda^2) \, ,
\]
so that $C=0$. The following identity is therefore established
(put $\alpha=1$):
\begin{equation}\label{KTdual}
\ln\chi_{KT}(\lambda)=
\tau\int\limits_0^\infty\frac{d\omega}{2\pi}\ln\left\{1+[1-
T_0(\omega)][n_L(1-n_R)(e^{-i\lambda}-1)+n_R(1-n_L)(e^{i\lambda}-1)]
\right\}
\end{equation}
The two equations (\ref{KT}) and (\ref{KTdual}) define the same
function, which means that the KT statistics, like the $g=1/2$
statistics, is reducible to the generic Levitov-Lesovik formula,
its relation to the non--interacting statistics
Eq.~(\ref{LLformula}) being
\begin{equation}\label{CBLL} \nonumber
\chi_{KT}(\lambda)= \chi_0 (-\lambda;V;\{1-T_0(\omega)\}) \;.
\end{equation}
Finally the explicit relation between the KT statistics and our
$g=1/2$ result is
\begin{equation}\label{iden}  \nonumber
\chi_{KT}(\lambda;V;\{T_0(\omega)\})=
\chi_{1/2}(-\lambda;V/2;\{1-T_0(\omega)\})\;,
\end{equation}
which is a direct consequence of the duality shown in
\cite{ourPRB}.

\section*{Appendix D}

The calculation starts as usual with the adiabatic potential [see
also Eq.~(\ref{firstU})],
\begin{eqnarray}                        \label{secondU}
 \frac{\partial}{\partial \lambda_-} \, {\cal U} (\lambda_\pm) =
 - \frac{1}{4} \, \int \, \frac{d \omega}{2 \pi} \left[ \gamma_+
 \left( \sin(\bar\lambda/4) G_{b \xi}^{--} + \cos(\bar\lambda/4) G_{b
 \eta}^{--} \right) + \gamma_-
 \left( \cos(\bar\lambda/4) G_{a \xi}^{--} - \sin(\bar\lambda/4) G_{a
 \eta}^{--} \right) \right] \, .
\end{eqnarray}
The most compact way to evaluate the inhomogeneous GFs entering
this expression is through their reduction to GFs involving only
the resonant level Majoranas $a$ and $b$. This is accomplished by
the following relations:
\begin{eqnarray}
 G_{b \eta}^{--} &=& i \gamma_+ \left[ D_{bb}^{--} \sin(\lambda_-/4)
 g_{\eta \eta}^{--} - D_{bb}^{-+} \sin(\lambda_+/4)
 g_{\eta \eta}^{+-}- D_{bb}^{--} \cos(\lambda_-/4)
 g_{\xi \eta}^{--} + D_{bb}^{-+} \cos(\lambda_+/4)
 g_{\xi \eta}^{+-}
  \right] \nonumber \\ \nonumber
&+& i \gamma_- \left[ D_{ba}^{--} \sin(\lambda_-/4)
 g_{\xi \eta}^{--} - D_{ba}^{-+} \sin(\lambda_+/4)
 g_{\xi \eta}^{+-} + D_{ba}^{--} \cos(\lambda_-/4)
 g_{\eta \eta}^{--} - D_{ba}^{-+} \cos(\lambda_+/4)
 g_{\eta \eta}^{+-}
  \right] \\ \nonumber
  G_{b \xi}^{--} &=& i \gamma_+ \left[ D_{bb}^{--} \sin(\lambda_-/4)
 g_{\eta \xi}^{--} - D_{bb}^{-+} \sin(\lambda_+/4)
 g_{\eta \xi}^{+-}- D_{bb}^{--} \cos(\lambda_-/4)
 g_{\xi \xi}^{--} + D_{bb}^{-+} \cos(\lambda_+/4)
 g_{\xi \xi}^{+-}
  \right] \nonumber \\ \nonumber
&+& i \gamma_- \left[ D_{ba}^{--} \sin(\lambda_-/4)
 g_{\xi \xi}^{--} - D_{ba}^{-+} \sin(\lambda_+/4)
 g_{\xi \xi}^{+-} + D_{ba}^{--} \cos(\lambda_-/4)
 g_{\eta \xi}^{--} - D_{ba}^{-+} \cos(\lambda_+/4)
 g_{\eta \xi}^{+-}
  \right] \\ \nonumber
 G_{a \eta}^{--} &=& i \gamma_+ \left[ D_{ab}^{--} \sin(\lambda_-/4)
 g_{\eta \eta}^{--} - D_{ab}^{-+} \sin(\lambda_+/4)
 g_{\eta \eta}^{+-}- D_{ab}^{--} \cos(\lambda_-/4)
 g_{\xi \eta}^{--} + D_{ab}^{-+} \cos(\lambda_+/4)
 g_{\xi \eta}^{+-}
  \right] \nonumber \\ \nonumber
&+& i \gamma_- \left[ D_{aa}^{--} \sin(\lambda_-/4)
 g_{\xi \eta}^{--} - D_{aa}^{-+} \sin(\lambda_+/4)
 g_{\xi \eta}^{+-} + D_{aa}^{--} \cos(\lambda_-/4)
 g_{\eta \eta}^{--} - D_{aa}^{-+} \cos(\lambda_+/4)
 g_{\eta \eta}^{+-}
  \right] \\ \nonumber
 G_{a \xi}^{--} &=& i \gamma_+ \left[ D_{ab}^{--} \sin(\lambda_-/4)
 g_{\eta \xi}^{--} - D_{ab}^{-+} \sin(\lambda_+/4)
 g_{\eta \xi}^{+-}- D_{ab}^{--} \cos(\lambda_-/4)
 g_{\xi \xi}^{--} + D_{ab}^{-+} \cos(\lambda_+/4)
 g_{\xi \xi}^{+-}
  \right] \nonumber \\ \nonumber
&+& i \gamma_- \left[ D_{aa}^{--} \sin(\lambda_-/4)
 g_{\xi \xi}^{--} - D_{aa}^{-+} \sin(\lambda_+/4)
 g_{\xi \xi}^{+-} + D_{aa}^{--} \cos(\lambda_-/4)
 g_{\eta \xi}^{--} - D_{aa}^{-+} \cos(\lambda_+/4)
 g_{\eta \xi}^{+-}
  \right]
\end{eqnarray}
Inserting these results into (\ref{secondU}) leads to
\begin{eqnarray}                \label{thirdU}
 \frac{\partial}{\partial \lambda_-} \, {\cal U} (\lambda_\pm) &=&
 - \frac{1}{4} \, \int \, \frac{d \omega}{2 \pi}
  \Big\{
  \gamma_+^2 \left[ D_{bb}^{--} g_{\eta \xi} - \cos(\bar\lambda/4)
  D_{bb}^{-+} g_{\eta \xi}^{+-} + \sin(\bar\lambda/4) D_{bb}^{-+}
  g_{\eta \eta}^{+-} \right]
  \nonumber \\
  &+&
  \gamma_-^2 \left[ D_{aa}^{--} g_{\eta \xi} - \cos(\bar\lambda/4)
  D_{aa}^{-+} g_{\eta \xi}^{+-} + \sin(\bar\lambda/4) D_{aa}^{-+}
  g_{\eta \eta}^{+-} \right]  \\ \nonumber
  &+& \gamma_+ \gamma_- \left[ (D_{ba}^{--} - D_{ab}^{--}) g_{\eta
  \eta}^{--} - \sin(\bar\lambda/4) (D_{ba}^{-+} - D_{ab}^{-+})g_{\eta
  \xi}^{+-} - \cos(\bar\lambda/4) (D_{ba}^{-+} - D_{ab}^{-+}) g_{\eta
  \eta}^{+-} \right] \Big\} \, .
\end{eqnarray}
Finally, the calculation of the composite $4 \times 4$ matrix
object
\begin{eqnarray}                    \nonumber
 \hat{\cal D} = \left[
  \begin{array}{cccc}
   D_{bb}^{--} &  D_{bb}^{-+} &  D_{ba}^{--} &  D_{ba}^{-+} \\
   D_{bb}^{+-} &  D_{bb}^{++} &  D_{ba}^{+-} &  D_{ba}^{++} \\
   D_{ab}^{--} &  D_{ab}^{-+} &  D_{aa}^{--} &  D_{aa}^{-+} \\
   D_{ab}^{+-} &  D_{ab}^{++} &  D_{aa}^{+-} &  D_{aa}^{++}
  \end{array}
  \right] \, ,
\end{eqnarray}
can be done by calculation of $\left[(\hat{\cal D}^{(0)})^{-1} -
\hat{\Sigma}_g \right]^{-1}$, where $\hat{\cal D}^{(0)} =
\mbox{diag}(1/\omega, -1/\omega, 1/\omega, -1/\omega)$ is the
corresponding matrix in the absence of the tunnelling couplings
and where the corresponding self-energy is given by
\begin{eqnarray}                \label{toinvert}   \nonumber
 \hat{\Sigma}_g =
 \left[ \begin{array}{cc}
          \hat{\Sigma}_{bb} &  \hat{\Sigma}_{ba} \\
          \hat{\Sigma}_{ab} &  \hat{\Sigma}_{aa} \\
        \end{array} \right] \, ,
\end{eqnarray}
and the components of this object are (we set $\Gamma_\pm =
\gamma_\pm^2/2$ and $\Gamma_\perp = \gamma_- \gamma_+ /2$),
\begin{eqnarray}             \nonumber
 \hat{\Sigma}_{bb} =
 \left[ \begin{array}{cc}
           \Delta^2/\omega + i \Gamma_+ (n_R + n_L - 1) & - i
           \Gamma_+ (n_L e^{i \bar\lambda/4} + n_R e^{- i \bar\lambda/4})  \\
           i \Gamma_+ \left[(1-n_R) e^{i \bar\lambda/4} + (1-n_L)e^{-
           i \bar\lambda/4} \right] & - \Delta^2/\omega + i \Gamma_+
           (n_R + n_L - 1)
        \end{array} \right] \, ,
\end{eqnarray}
\begin{eqnarray}          \nonumber
 \hat{\Sigma}_{ba} =
  \Gamma_\perp \left[ \begin{array}{cc}
            n_R - n_L &   n_L e^{i
           \bar\lambda/4} - n_R e^{-i \bar\lambda/4} \\
            (1-n_R) e^{i
           \bar\lambda/4} - (1-n_L) e^{- i \bar\lambda/4} &
            n_R - n_L
        \end{array} \right] \, ,
\end{eqnarray}
\begin{eqnarray}    \nonumber
 \hat{\Sigma}_{ab} =
\Gamma_\perp \left[ \begin{array}{cc}
            n_L - n_R &  - n_L e^{i
           \bar\lambda/4} + n_R e^{-i \bar\lambda/4} \\
            - (1-n_R) e^{i \bar\lambda/4} + (1-n_L)
           e^{- i \bar\lambda/4} & n_L - n_R
        \end{array} \right] \, ,
\end{eqnarray}
\begin{eqnarray}   \nonumber
 \hat{\Sigma}_{aa} =
 \left[ \begin{array}{cc}
           \Delta^2/\omega +
           i \Gamma_- (n_R + n_L - 1) & - i \Gamma_- (n_L e^{i
           \bar\lambda/4} + n_R e^{-i \bar\lambda/4}) \\ i
           \Gamma_- \left[ (1-n_R) e^{i \bar\lambda/4} + (1-n_L) e^{-
           i \bar\lambda/4} \right] & - \Delta^2/\omega + i \Gamma_-
           (n_R + n_L - 1)
        \end{array} \right] \, .
\end{eqnarray}
Unsurprisingly, an inversion of the matrix $(\hat{\cal
D}^{(0)})^{-1} - \hat{\Sigma}_g$ and substitution of the result
into (\ref{thirdU}) leads to exactly the same expression for the
statistics (\ref{12stat}) up to the transmission coefficient,
which is now more involved and which is the same as calculated
previously in the context of the non-linear $I-V$ of the same
set-up \cite{ourPRL}.

\end{widetext}

\bibliography{fcsPRB}

\end{document}